\pdfoutput=1 
\documentclass[11pt]{article}
\usepackage{mydef2col}
\usepackage{kantlipsum,widetext}
\usepackage[utf8]{inputenc}
\usepackage[T1]{fontenc}
\usepackage[autostyle=true]{csquotes}
\usepackage{regexpatch}
\usepackage{cprotect}

\usepackage[colorinlistoftodos,prependcaption,textsize=tiny]{todonotes}
\makeatletter
\xpatchcmd{\@todo}{\setkeys{todonotes}{#1}}{\setkeys{todonotes}{inline,#1}}{}{}
\makeatother

\allowdisplaybreaks[4]
\graphicspath{{./Graphs/}}
\usepackage{tikz}
\usetikzlibrary{calc,decorations.markings}
\usetikzlibrary{arrows,patterns,positioning}

\usepackage[comma, sort&compress]{natbib}
\setcitestyle{square}

\def\calK{{\cal K}}

\def\Ito{It\^{o}'s }

\def\erf{\operatorname{Erf}}
\def\erfc{\operatorname{Erfc}}
\def\erfi{\operatorname{Erfi}}

\def\JnuM{{J_{|\nu|}}}
\def\Jnum{{J_{-\nu}}}

\def\tth{{\tilde{\theta}}}
\def\br{{\bar{r}}}
\def\calI{{\cal I}}
\def\calJ{{\cal J}}

\newcommand{\iu}{\mathrm{i}\mkern1mu}

\newcommand{\Ind}{\mathbf{1}}

 \title{American options in time-dependent one-factor models: \\[0.2em]\smaller{}Semi-analytic pricing, numerical methods and ML support}
\shorttitle{American options in time-dependent one-factor models}
\author{
\authorstyle{Andrey Itkin\textsuperscript{1, 2}
and Dmitry Muravey\textsuperscript{2}
}
\newline\newline
\textsuperscript{1}
\institution{Tandon School of Engineering, New York University, USA,} \\
\textsuperscript{2}
\institution{Abu Dhabi Investment Authorities, UAE.}
}

\date{\today}

\begin{document}

\maketitle

{\bf Abstract:} Semi-analytical pricing of American options in a time-dependent Ornstein-Uhlenbeck model was presented in \citep{CarrItkin2020jd}. It was shown that to obtain these prices one needs to solve (numerically) a nonlinear Volterra integral equation of the second kind to find the exercise boundary (which is a function of the time only). Once this is done, the option prices follow. It was also shown that computationally this method is as efficient as the forward finite difference solver while providing better accuracy and stability. Later this approach called "the Generalized Integral transform" method has been significantly extended to various time-dependent one factor, \citep{ItkinLiptonMuraveyBook}, and stochastic volatility \citep{CarrItkinMuraveyHeston, ItkinMuraveySabrJD} models as applied to pricing barrier options. However, for American options, despite possible, this was not explicitly reported anywhere. In this paper our goal is to fill this gap and also discuss which numerical method could be efficient to solve the corresponding Volterra equations, also including machine learning.

%%%%%%%%%%%%%%%%%%%%%%%%%%%%%%%%%%%%%%%%%%%%%%%%%%%%%%%%%%%%
\vspace{0.5in}

\section*{Introduction}

An {\it American} option is a derivative contract (option) that can be exercised at any time during its life, i.e. prior to and including its maturity date, see, e.g., \citep{Detemple2006, hull97} and references therein among others. This flexibility brings some advantage to the option holder as compared with holding the corresponding European option which can only be exercised at maturity. Nowadays, most stocks and exchange-traded funds have American-style options (while European-style options. are traded for, e.g., equity indices).

Another example of an early exercise feature involved into the contract terms are {\it real} options. By definition, real options differ from their financial counterparts since they involve physical assets as the underlying, rather than financial security for normal options, and are not exchangeable as securities. It is important that real options do not refer to a derivative financial instrument, such as call and put options contracts, which give the holder the right to buy or sell an underlying asset, respectively. Instead, real options are opportunities that a business may or may not take advantage of or realize.

Pricing of American (or Bermudan) options is more sophisticated as compared with the European ones since it requires solution of a linear complementary problem that normally can be done only numerically. Various efficient numerical methods have been proposed for doing that, see \citep{SahaliaCarr1997, IkonenToivanen2007, Fasshauer2, Kohler2010, NBR2019, ItkinBook} among many others. For real options some valuation models also use terminology from derivatives markets wherein the strike price corresponds to non-recoverable costs involved with the project. Similarly, the expiration date of an options contract could be substituted with the timeframe within which the business decision should be made. Real options must also consider the risk involved, and it too could be assigned a value similar to volatility. Accordingly, most of the classical applications of real options analysis involve vanilla American options as the case of the option to postpone a project, or to abandon it, see \citep{Brigatti2015} and references therein among the others. Pricing of real options usually requires some flavor of the Monte Carlo method.

As mentioned in \citep{CarrItkin2020jd}, there exists another approach constructed based on the Generalized Integral Transform (GIT) method which utilizes a notion of the exercise boundary. By definition, the boundary $S_B(t)$ is defined in such a way, that, e.g., for the American Call option $P_A(S,t)$ at $S \le S_B(t)$ it is always optimal to exercise the option, therefore  $P_A(S,t) = K - S$. For the complementary domain $S > S_B(t)$ the earlier exercise is not optimal, and in this domain $P_A(S,t)$ solves the corresponding Kolmogorov equation. This domain is called the continuation (holding) region. However, $S_B(t)$ is not known in advance, and instead we only know the price of the American option at the boundary. For instance, for the American Put we have $P_A(S_B(t),t)  = K - S_B(t)$, and for the American Call - $C_A(S_B(t),t)  = S_B(t) - K$. A typical shape of the exercise boundary for the Call option is presented in Fig.~\ref{EB}.

Therefore, instead of calculating American option prices directly, one can find an explicit location of the option exercise boundary. This approach was advocated, e.g., in \citep{Andersen2016} for the Black-Scholes model with constant coefficients. It was shown that $S_B(t)$ solves an integral (Volterra) equation which can be solved numerically. The proposed numerical scheme can be implemented straightforward, and it converges at a speed several orders of magnitude faster than the other (previously mentioned) approaches.

As shown in \citep{CarrItkin2020jd}, finding the exercise boundary is almost equivalent to pricing a barrier option under the same model where the boundary is time dependent. Let us denote $S_t$ the relative underlying price at the time $t > 0$, , and $h(t)$ the time dependent barrier price, say for the Up-and-Out barrier option. The important difference between pricing American and barrier options is as follows

\begin{itemize}
\item For the barrier option pricing problem the moving boundary (the time-dependent barrier) is known. But the Option Delta $\fp{P}{S}$ at the boundary $S = h(t)$ is not, and should be found by solving the Volterra integral equation, in more detail see \citep{ItkinLiptonMuraveyBook}.

\item For the American option pricing problem the moving boundary is not known. However, the option Delta $\fp{P_A}{S}$ ($\fp{P_A}{S}$) at the boundary $S = h(t)$ is known (it follows from the exercise conditions $\fp{C_A}{S}|_{S = S_B(t)} = 1$ and $\fp{P_A}{S}|_{S = S_B(t)} = -1$. Also the boundary condition for the American Call and Put at the exercise boundary (the moving boundary) differs from that for the Up-and-Out barrier option, namely: it is $C_A(S_B(t),t)) = S_B(t) - K$ for the Call, and $P_A(S_B(t),t)) = K - S_B(t)$ for the Put.

\item In accordance to the above items the Volterra integral equation of the second kind remains the same for both problems. However, for the barrier options it is linear in the unknown option Delta at the boundary, while for the American options is non-linear for the unknown exercise (moving) boundary.
\end{itemize}

Because of the similarity of these two problems, it turns out that the American option problem can be solved if a) the solution for the continuation region (for the European option) is known, and b) the exercise boundary is found by using the approach proposed in \citep{CarrItkin2020jd, ItkinLiptonMuraveyBook}. However, due to the reported differences numerical methods for solving the corresponding Volterra equations could vary.

Note, that efficient pricing of American options remains to be an important problem for decades. For instance, nowadays, the modern research has a significant focus on {\it real options}  where the American feature is an embedded part of the option. By definition, a real option is an economically valuable right to make or else abandon some choice that is available to the managers of a company, often concerning business projects or investment opportunities. It is referred to as “real” because it typically references projects involving a tangible asset (such as machinery, land, and buildings, as well as inventory), instead of a financial instrument, see \citep{Hayes2021, Damodaran2008} among others.

As shown in \citep{ItkinLiptonMuraveyBook} for various barrier options our approach is an important alternative to the existing methods providing various additional advantages. For instance, in \citep{ItkinLiptonMuraveyMulti} the authors emphasize that, traditional finite-difference (FD) methods provide only the values of the unknown function on the grid nodes in space, and at intermediate points they can be found only by interpolation. In contrast, the GIT method derives an analytic representation of the solution at any $x$. Second, the Greeks, i.e., derivatives of the solution, can be expressed semi-analytically by differentiating the solution with respect to $x$ or any necessary parameter of the model and then performing numerical integration. For the FD method, the Greeks can be found only numerically. Moreover, to compute, e.g. the option Vega, a new run of the FD method is required, while for the GIT method all Greeks can be calculated in one sweep.

Theoretically, same should be true for American options as well. However, yet these results were not presented explicitly in the literature. Therefore, in this paper we fulfill this gap for several time-dependent one-factor models by first, presenting the corresponding integral Volterra equation for $S_B(t)$ and discussing it in detail, and second, discussing various numerical method that could be efficient for solving this class of integral equations.

Same approach could potentially be applied to stochastic volatility models as this was done in \citep{ItkinMuraveySabrJD,
CarrItkinMuraveyHeston}. We intend to present these results elsewhere.

The rest of the paper is organized as follows. In Section~\ref{SecHeat} we consider examples of one-factor time-dependent models where the pricing PDE can be reduced to the heat equation, in particular, a time-dependent  Ornstein-Uhlenbeck (OU) model (Section~\ref{tdOU}), a time-dependent Hull-While model (Section~\ref{tdHW}), and a time-dependent Verhulst model, (Section~\ref{secVerhulst}). For each model we derive nonlinear integral Volterra equations w.r.t. the unknown exercise boundary $S_B(t)$. Then, in Section~\ref{interCom} we provide some general comments to the proposed approach and also present an example how the obtained integral equations could be solved by using a simple numerical algorithm. In Section~\ref{secBess} a similar scheme is developed for models where the pricing PDE can be reduced to the Bessel equation. In particular, we consider the CEV model (Section~\ref{SecBO}) and the CIR model (Section~\ref{sCIR}), Again, a nonlinear integral Volterra equation w.r.t. the unknown exercise boundary $S_B(t)$ is derived for each model (Section~\ref{nonLinVoltCEV}). Section~\ref{ML} describes various numerical methods that can be used to solve these integral equations, also including those using Machine Learning (ML) methods. The final Section concludes.

\section{One-factor models that can be reduced to the heat equation} \label{SecHeat}

In this section we consider two typical examples of one-factor time-dependent models such that the American option pricing can be reduced to solving the heat equation at the time-dependent spatial domain. We chose pricing of the American Call option written on a stock which follows the time-dependent OU model, and pricing of the American Put option written on a zero-coupon bond where the underlying interest rate follows the time-dependent Hull-White model.

\subsection{The time-dependent Ornstein-Uhlenbeck model} \label{tdOU}

Pricing American options in the time-dependent OU model by using the GIT method was first presented in \citep{CarrItkin2020jd}. The authors derived a nonlinear Fredholm equation of the first kind with respect to the unknown exercise boundary function $y(t)$ which be solved numerically. Then an analytic representation of the American option price is obtained. Below we shortly describe the model and the approach in use (since this would be helpful for the remaining Sections), and also instead of the Fredholm equation derive an analogues nonlinear Volterra equation of the second kind which is more stable when solved numerically.

To remind, in the time-dependent OU model the underlying spot price $S_t$ follows the stochastic differential equation (SDE), \citep{andersen2010interest, Thomson2016}

\begin{equation} \label{OU1}
d S_t = [r(t) - q(t)] S_t dt + \sigma(t)dW_t,
\end{equation}
\noindent where $r(t)$ is the deterministic short interest rate, $q(t)$ is the continuous dividend yield, $\sigma(t)$ is the volatility and $W_t$ is the Brownian motion process. This model is also known in the financial literature as the Bachelier model. One can think about $S_t$, e.g., as the stock price or the price of some commodity asset. While in the Bachelier model the underlying value could become negative which is not desirable  for the stock price, this is fine for commodities under the modern market conditions when the oil prices have been several times observed to be negative, see e.g., \citep{CME}. For the sake of certainty, below we will reference $S_t$ as the stock price.

In \eqref{OU1} we don't specify the explicit form of $r(t), q(t), \sigma(t)$ but assume that they are known as a differentiable functions of time $t \in [0,\infty)$. The case of discrete dividends is discussed in \citep{CarrItkin2020jd}.

\begin{figure}
\centering
\includegraphics[totalheight=2.7in]{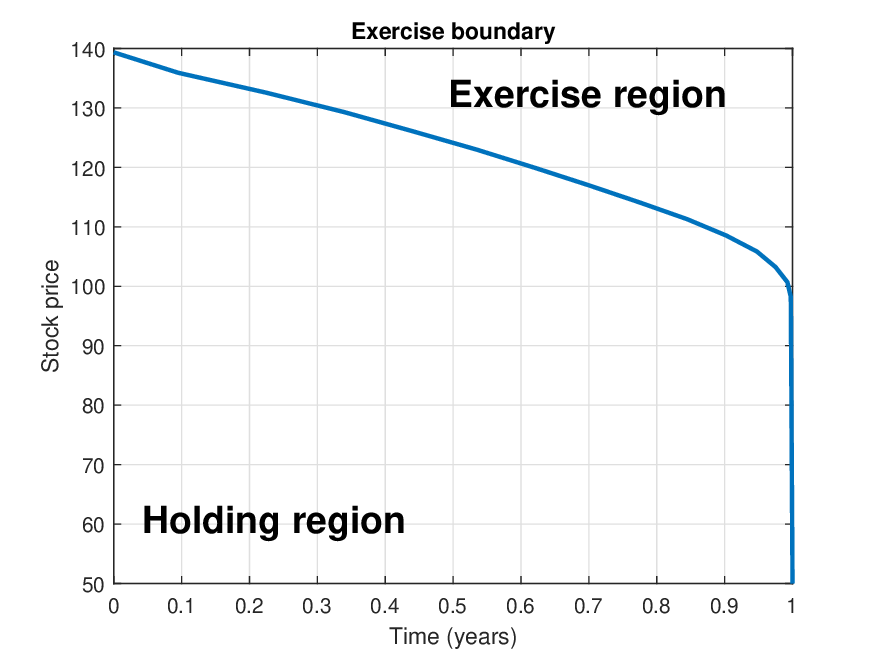}
\caption{Typical exercise boundary for the American Call option under the Black-Scholes model with $K = 50, r = 0.2, q = 0.1, \sigma = 0.5, T = 1$.}
\label{EB}
\end{figure}

Let's consider an American option (Call) written on the underlying process $S_t$ in \eqref{OU1} with the strike price $K$ which can be exercised at any time $0 < t \le T$, $T$ the time to maturity. It is known, that there exists a function $S_B(t)$ - the optimal (exercise) boundary, which splits the whole domain $\Omega: S \in [0, \infty) \times t \in [0,T]$ into two regions (see an example in Fig.~\ref{EB}): the continuation (holding) region where it is not optimal to exercise the option, and the exercise region with the opposite optimality\footnote{It seems to be an interesting question whether the exercise boundary $S_B(t)$ for the American Call should be convex or not, and also increase or decrease with time. For instance, in \citep{Kwok2022}, in Fig~5.2 the boundary is increasing with time and concave. Also, in \citep{Chen2007} the authors argue that  non-convex free boundary seems to be the most likely case; e.g., on a dividend–paying asset, numerical simulations by J. Detemple suggest that the early exercise boundary may not be convex for all choices of the parameters.}

By a standard argument, \citep{ContVolchkova2005, klebaner2005} the Call option price $C(S,t)$ in the continuation area solves a parabolic partial differential equation (PDE)
\begin{equation} \label{PDE}
\fp{C}{t} + \dfrac{1}{2}\sigma^2(t) \sop{C}{S} +  [r(t) - q(t)] S \fp{C}{S} = r(t) C.
\end{equation}
\noindent subject to the terminal condition at the option maturity $t=T$
\begin{equation} \label{tc0}
C(T,S) = (S-K)^+,
\end{equation}
\noindent and the boundary conditions
\begin{equation} \label{bc0}
C(t,0) = 0, \qquad \Delta(t,S_B(t)) = 1, \qquad \Delta(t,S) = \fp{C(t,S)}{S}.
\end{equation}

Note, that for the arithmetic Brownian Motion process the domain of definition is $S \in (-\infty,\infty)$, however here we move the boundary condition from minus infinity to zero, see \citep{ItkinMuravey2020r} about rigorous boundary conditions for this problem.  This is because in practice we can control the left boundary to make the probability of $S$ dropping below 0 to be low.

Further let us assume that the exercise boundary is somehow known. Then $C(S,t)$ can be found by solving \eqref{PDE} subject to the terminal condition in \eqref{tc0} and the boundary conditions
\begin{equation} \label{bc1}
C(t,0) = 0, \qquad C(t,S_B(t)) = S_B(t) - K.
\end{equation}
In other words, the American option price $C(S,t)$ in this region concises with the price of a double barrier option with the same maturity $T$, strike $K$, the constant lower barrier $L = 0$ with no rebate at hit, and the time-dependent upper barrier $H = S_B(t)$. When the upper barrier is reached, the option is terminated, and the option holder gets a rebate at hit in the amount $S_B(t) - K$.

Obviously, in the exercise region the undiscounted Call option value is $C(t,S) = (S-K)^+ \mathds{1}_{S \ge S_B(t)}$, where $\mathds{1}_{x}$ is the indicator function.

Below across this entire paper and for all presented models we assume that the pricing problem for the corresponding barrier option can be solved semi-analytically as this is shown in \citep{ItkinLiptonMuraveyBook, ItkinMuraveySabrJD, CarrItkinMuraveyHeston}. Accordingly, our main focus here will be on i) deriving an integral equation for the exercise boundary $S_B(t)$ for each model considered in the paper, and ii) developing efficient methods (semi-analytical or numerical)  for solving these integral equations.

Another comment should be made about using the GIT technique for solving the problems in this paper. Since all the problems have a vanishing terminal condition, a method of Heat potentials (HP) can also be used for doing so, \citep{ItkinLiptonMuraveyBook}. However, the HP method would be more expensive in this case as compared with the GIT. That is, because in the HP method we first express the solution via the heat potential density $\Omega(\tau)$ and obtain an integral Volterra equation which $\Omega$ solves. Then, another integral expression of the gradient of the solution at the moving boundary is obtained, in more detail \citep{ItkinLiptonMuraveyBook}, Chapter~8. Thus, instead of a single Volterra equation as for the GIT, here we need to solve a system of two Volterra equations, hence this method is less efficient.

\subsubsection{Nonlinear integral Volterra equation for $S_B(t)$} \label{voltOU}

In \citep{CarrItkin2020jd} the PDE in \eqref{PDE} by a series of transformations  has been reduced to the heat equation \begin{equation} \label{Heat}
\fp{u}{\tau} = \sop{u}{x}, \qquad (\tau , x) \in \mathbb{R}_+ \times [0, y(\tau)] ,  \qquad
y(\tau) = S_B(t(\tau)) g(t(\tau)), \\
\end{equation}
\noindent which should be solved subject to the terminal condition
\begin{equation} \label{tcA}
u(0,x) = \left(x g(T) - K\right)^+ e^{-f(T,x)},
\end{equation}
\noindent and the boundary conditions
\begin{align} \label{bcA}
u(\tau,0) &= 0, \qquad u(\tau, y(\tau)) = y(\tau) - K.
\end{align}

Here, first a change of independent variables was made
\begin{align} \label{indV}
x &= S g(t), \qquad g(t) = e^{\int_T^t w(s) ds}, \qquad
\tau(t) = \frac{1}{2} \int_t^T \sigma^2(s) e^{2 \int_0^s w(m) \, dm} \, ds,
\end{align}
\noindent (so,  $y(\tau) = S_B(t(\tau)) g(t(\tau))$), where the function $w(t)$ solves the Riccati equation
\begin{equation} \label{Ric}
w'(t) = 2 (r(t) - q(t)) \frac{\sigma'(t)}{\sigma(t)} - \left[ (r(t)-q(t))^2 + r'(t) - q'(t) \right] + w(t)^2  + 2 w(t) \frac{\sigma '(t)}{\sigma (t)},
\end{equation}
\noindent followed by a change of the dependent variable\footnote{To remind, in the below expression $\sigma(t)$ is the normal volatility, therefore the ratio $x^2/\sigma^2(t)$ has dimensionality $1/t$, hence $f(x,t)$ is dimensionless.}
\begin{align} \label{fDef}
u(\tau, x) &= C(S,t(\tau)) e^{-f(t(\tau),x)}, \\
f(t,x) &= \frac{1}{2} \log \left(\frac{g(t)}{g(T)}\right) + \frac{1}{2} \int_T^t [3 r(s)-q(s)] ds  - \frac{g'(t)+g(t) (r(t)-q(t))}{2 g(t)^3 \sigma (t)^2} x^2.  \nonumber
\end{align}

The Riccati equation in \eqref{Ric} cannot be solved analytically for arbitrary functions $r(t), q(t), \sigma(t)$, but can be efficiently solved numerically. Also, in some cases it can be solved in closed form, see \citep{CarrItkin2020jd}.

Since the holding region is defined as $(\tau, x) \in [0, \tau(0)] \times [0, y(\tau)]$ (this is the area under the exercise boundary curve in Fig.~\ref{EB}) and $g(0) = 1, \ y(0) = K$ \footnote{The asymptotic value of the exercise boundary at $\tau \to 0^+$ for an American Call option is, \citep{Kwok2022}
\begin{equation}
y(0) =
\begin{cases}
K, & r(0) \ge q(0), \\
\frac{r(0)}{q(0)} K, & r(0) > q(0).
\end{cases}
\end{equation}
Hence, for $q(0) < r(0)$, there is a jump of $y(\tau)$ at $\tau = 0$.
}
, the terminal condition in \eqref{tcA} becomes homogeneous, i.e.
\begin{equation} \label{tcFin}
u(0,x) = 0.
\end{equation}
Thus, the \eqref{Heat} is a PDE with the homogeneous initial condition and the boundary condition at $x=0$ and an inhomogeneous boundary condition at $x = y(\tau)$. It turns out that this problem has been already solved in \citep{ItkinMuraveyDBFMF} where pricing problem of double barrier options was investigated in detail. In terms of that paper, our problem in  \eqref{Heat} can be treated as a pricing problem for the double barrier option with the lower barrier $y_B(\tau) = 0$ and the upper barrier $z_B(\tau) = y(\tau)$ and also rebates paid at hit. The boundary conditions (rebates) at these barriers accordingly are
\begin{equation} \label{bcOU2}
f^-(\tau) \equiv u(\tau,y_B(\tau)) = 0, \qquad f^+(\tau) \equiv u(\tau, z_B(\tau)) = y(\tau) - K.
\end{equation}

It is worth mentioning that this problem by a change ov variables $U(\tau,x) = u(\tau, x) - \frac{x }{y(\tau)}f^+(\tau)$ can be transformed to a similar problem, but with all terminal and boundary conditions to be homogeneous
\begin{align} \label{wEq}
\fp{U}{\tau} &= \sop{U}{x} + \lambda(\tau,x), \qquad \lambda(\tau,x) = -x \frac{\partial}{\partial \tau} \left( \frac{f^+(\tau)}{y(\tau)}\right), \\
U(0,x) &=U(\tau,0) = U(\tau, y(\tau)) = 0. \nonumber
\end{align}
Then the Duhamel's principle can be applied to solve this problem if the solution of the same problem but with no source term is known\footnote{We didn't find in the literature any reference to using the Duhamel's formula for problems with moving boundaries. Therefore, in Appendix~\ref{app1} we directly derive this formula from first principles for the domain $x \in [0, y(\tau)]$. Also, in the next Section we solve a similar problem for the domain $x \in [y(\tau),\infty)$ directly. Using the result obtained, we formulate how the Duhamel's principle looks in both cases.}.

The solution of \citep{ItkinMuraveyDBFMF} applied to our problem in \eqref{Heat} reads
\begin{align}  \label{uFourier}
u(\tau, x) &= \int_0^\tau \Psi(s, y(s)) B_0(x,\tau \,|\, y(s), s) +\int_0^\tau \int_{0}^{y(s)} \lambda(s, \xi) B_0(x,\tau \,|\, \xi, s)  d\xi ds + \frac{x}{y(\tau)} f^+(\tau), \nonumber \\
\Psi(\tau, y(\tau)) &= - \frac{y(\tau)-K}{y(\tau)} + \fp{u(\tau, x)}{x} \Bigg|_{x = y(\tau)}, \\
B_0(x,\tau \,|\, \xi, s) &= \sum_{n=-\infty}^{\infty} \Upsilon_n (x, \tau \,|\, \xi, s), \qquad
\Upsilon_n(x, \tau \,|\, \xi, s) = \frac{1}{2\sqrt{\pi (\tau - s)}}\left[e^{-\frac{(2n y(\tau)  + x - \xi)^2}{4 (\tau - s)}} - e^{-\frac{(2n y(\tau)  + x +  \xi)^2}{4 (\tau - s)}} \right].  \nonumber
\end{align}
In what follows we will also need the following function
\begin{align}
B_1(x,\tau \,|\, \xi, s) &= \sum_{n=-\infty}^{\infty} \Lambda_n(x, \tau  \,|\, \xi, s), \\
\Lambda_n(x, \tau \,|\, \xi, s) &= \frac{\partial }{\partial \xi} \Upsilon_n(x, \tau \,|\, \xi, s) = \frac{x - \xi + 2n y(\tau)}{4 \sqrt{\pi (\tau -s)^3}} e^{-\frac{(2n y(\tau)  + x  - \xi)^2}{4 (\tau - s)}} + \frac{x + \xi + 2n y(\tau) }{4 \sqrt{\pi (\tau -s)^3}} e^{-\frac{(2n y(\tau)  + x +  \xi)^2}{4 (\tau - s)}}. \nonumber
\end{align}

Note that the Fourier series in these expressions usually converge rapidly when $n$ grows.  Similarly, taking the derivative of this series on $x$ provides a convenient way of calculating the corresponding derivative $\fp{u(\tau,x)}{x}$, \citep{NIST:DLMF}.

An alternative representation can be obtained in terms of the Jacobi theta functions of the third kind $\theta_3 (z,\omega)$, \citep{mumford1983tata}
\begin{align}  \label{uTheta}
B_0(x,\tau \,|\, \xi, s) &= \frac{1}{2 y(\tau)}\left[ \theta_3(\phi_-(x,\xi,\tau), \omega_2(s, \tau)) - \theta_3(\phi_+(x,\xi,\tau),\omega_2(s, \tau)) \right], \\
B_1(x,\tau \,|\, \xi, s) &= - \frac{\pi}{4 y(\tau)^2} \left[ \theta'_3(\phi_-(x,\xi,\tau), \omega_2(s, \tau)) + \theta'_3(\phi_+(x,\xi,\tau),\omega_2(s, \tau)) \right]. \nonumber
\end{align}
Here
\begin{align} \label{notation}
\theta_3 (z,\omega) &= 1 + 2 \sum_{n=1}^{\infty} \omega^{n^2}\cos(2 n z), \quad \omega_1(s, \tau) = e^{ - \left( \frac{\pi \sqrt{s}}{y(\tau)}\right)^2}, \quad \omega_2(s,\tau) = e^{ \left( \frac{\pi \sqrt{s-\tau}}{y(\tau)}\right)^2}, \\
\phi_-(x,z,\tau) &= \frac{\pi (x-z)}{2 y(\tau)},  \quad \phi_+(x,z,\tau) = \frac{\pi (x+z)}{2 y(\tau)}. \nonumber
\end{align}
A well-behaved theta function must have parameter $|\omega| < 1$, \citep{mumford1983tata}. This condition holds for any $\tau > 0$. Also, $\theta'_3(x, \omega)$ is the first derivative of the Jacobi Theta function on $x$ (e.g., \verb|EllipticThetaPrime| in Wolfram Mathematica).

The formulas \eqref{uFourier} and \eqref{uTheta} are complementary. Since the exponents in \eqref{uTheta} are proportional to the difference $\tau - s$, the Fourier series \eqref{uTheta} converge fast if $\tau - s$ is large. Contrary, the exponents in \eqref{uFourier} are inversely proportional to $\tau - s$. Therefore, the series \eqref{uFourier} converge fast if $\tau - s$ is small. As mentioned in \citep{ItkinMuraveyDBFMF}, this situation is well investigated for the heat equation with constant coefficients. There exist two representation of the solution: one - obtained by using the method of images, and the other one - by the Fourier series. Despite both solutions are equal as the infinite series, their convergence properties are different, \citep{Lipton2001}.

It is easy to see that by definition, $\Psi(\tau, y(\tau))$ is the Call option Delta (shifted by $1 - K/y(\tau)$ ) at the exercise boundary but expressed in new dependent and independent variables. Since in the original variables it is given by \eqref{bc0}, it can be explicitly expressed in the new variables as well to obtain
\begin{align} \label{uEB}
u(\tau,y(\tau)) &= \left[ \frac{y(\tau)}{g(\tau)} - K\right] e^{-f(t(\tau), y(\tau))}, \\
\Psi(\tau, y(\tau)) &= - \frac{y(\tau)-K}{y(\tau)} + \frac{e^{- f(t(\tau), y(\tau))}}{g(\tau)}\left[1 + y(\tau) \left(y(\tau) - K g(\tau) \right) a(t(\tau))\right], \nonumber \\
a(t(\tau)) &= \frac{g'(t) + g(t) (r(t)-q(t))}{g^3(t) \sigma (t)^2}.  \nonumber
\end{align}
Thus, \eqref{uFourier}, \eqref{uTheta} provides a semi-analytic expression of the American Call option price if the exercise boundary $y(\tau)$ is known.

If $y(\tau)$ is not known, it can be found by solving a Volterra integral equation. At the first glance, it can be obtained by setting $x = y(\tau)$ in \eqref{uFourier} and substitute the values of $u(\tau,y(\tau)), \Psi(\tau, y(\tau))$ from \eqref{uEB} into it. However, it can be verified that in this case $B_0(y(\tau),\tau \,|\, y(s), s) = B_1(y(\tau),\tau \,|\, y(s), s) = 0$, i.e., \eqref{uFourier} reduces to the correct boundary condition \eqref{bcOU2} at the moving boundary.
Therefore, the right way of finding $y(\tau)$ is similar to how the Volterra equation for the unknown function $\Psi(\tau)$ is derived when using the GIT method. Both parts of \eqref{uFourier} can be first, differentiated by $x$, and then $x = y(\tau)$ should be substituted into the result. This yields, \citep{ItkinMuraveyDBFMF}
\begin{align} \label{finVolterra}
\Psi(\tau, y(\tau)) &=  - \int_0^\tau \Psi(s, y(s)) \left[ \frac{y(\tau) - y(s)}{2\sqrt{\pi (\tau -s)^3}} e^{- \frac{(y(\tau) - y(s))^2}{4(\tau -s)}} - \upsilon^+_{0}(\tau \,|\,s, y(s)) \right] ds \nonumber \\
&+ \int_0^\tau \int_{0}^{y(s)} \lambda(s, \xi) B_1(y(\tau),\tau \,|\, \xi, s)  d\xi ds + \frac{1}{y(\tau)} f^+(\tau),
\end{align}
\noindent where
\begin{align}
%\eta^+(\tau \,|\, \xi, s) &= \frac{1}{\sqrt{\pi (\tau -s)}} \sum_{n = -\infty}^\infty e^{-\frac{(- \xi + (2n +1) y(\tau))^2}{4(\tau -s)}} - \frac{1}{\sqrt{\pi (\tau -s)}}, \nonumber \\
\upsilon^+_0(\tau \,|\, s, \xi) &= \sum_{\substack{n = -\infty \\ n \neq 0}}^{\infty}  \frac{\xi - (2 n+1) y(\tau)}{2 \sqrt{\pi (\tau - s)^3}} e^{-\frac{(- \xi + (2 n+1) y(\tau))^2}{4 (\tau -s)}}.
\end{align}
An alternative representation can be obtained in terms of the Jacobi theta function $\theta_3(z, \omega)$
\begin{align} \label{eta_def_Theta}
%\eta^+(\tau \,|\, \xi, s) &= \frac{1}{y(\tau)} \theta_3\left( \phi_-(y(\tau),\xi, \tau), \omega_2(s, \tau) \right)
%- \frac{1}{\sqrt{\pi  (\tau -s)}}, \\
\upsilon^+_0(\tau \,|\, s, \xi) &= \frac{\pi}{2 y(\tau)^2} \theta'_3\left( \phi_-(y(\tau),\xi, \tau), \omega_2(s, \tau) \right) + \frac{y(\tau )-\xi}{2 \sqrt{\pi (\tau -s)^3}} e^{-\frac{(\xi -y(\tau ))^2}{4 (\tau-s)}}.
\end{align}
It can be checked that $f^+(0) = 0$.
%The first and the last integral in \eqref{finVolterra} can also be integrated by parts to yield
%\begin{align}
%\int_0^\tau \Bigg\{ \frac{f^+(s) - f^+(\tau)}{2 \sqrt{\pi (\tau - s)^3}} ds &= \frac{f^+(\tau)}{\sqrt{\pi \tau}}
% - \int_0^\tau  \frac{d f^+(s)}{ds} \frac{ds}{\sqrt{\pi (\tau - s)}}
%\\
%\int_0^\tau  f^+(s) \frac{d}{ds} \left(\eta^+(\tau \,|\, y(s), s) \right) ds &= - \int_0^\tau  \frac{df^+(s)}{dy(s)} \eta^+(\tau \,|\, y(s), s) ds. \nonumber .
%\end{align}
Substituting these formulae into \eqref{finVolterra} and taking into account $f^+(\tau) = y(\tau) - K$ we arrive at the following replacement for \eqref{finVolterra}
\begin{align} \label{finVolterra1}
\Psi(\tau, y(\tau)) &=-\int_0^\tau \Psi(s,y(s)) \left[ \frac{y(\tau) - y(s)}{2\sqrt{\pi (\tau -s)^3}} e^{- \frac{(y(\tau) - y(s))^2}{4(\tau -s)}} - \upsilon^+_{0}(\tau \,|\,s, y(s)) \right] ds \\
%&+\int_0^\tau y'(s) \left(\frac{1}{\sqrt{\pi (\tau -s )}}+ \eta^+(\tau | y(s), s)\right) ds
&+ \int_0^\tau \int_{0}^{y(s)} \lambda(s, \xi) B_1(y(\tau),\tau \,|\, \xi, s)  d\xi ds + \frac{1}{y(\tau)} f^+(\tau), \nonumber
\end{align}
\noindent which by substituting the definitions from \eqref{eta_def_Theta} takes a form
\begin{align} \label{finVolterra2}
\Psi(\tau, y(\tau)) &=  \frac{1}{y(\tau)} f^+(\tau) + \frac{\pi}{2 y(\tau)^2} \int_0^\tau \Psi(s,y(s))  \theta'_3\left( \phi_-(y(\tau),y(s), \tau), \omega_2(s, \tau) \right) ds \\
%&+ \frac{1}{y(\tau)} \int_0^\tau y'(s) \theta_3\left( \phi_-(y(\tau), y(s), \tau), \omega_2(s, \tau) \right) ds
&+ \int_0^\tau \int_{0}^{y(s)} \lambda(s, \xi) B_1(y(\tau),\tau \,|\, \xi, s)  d\xi ds. \nonumber
\end{align}
Since $\lambda(\tau,x)$ in \eqref{wEq} is a linear function of $x$, the inner integral in the last term can be further simplified to yield
\begin{align}
\int_{0}^{y(s)} & \lambda(s, \xi) B_1(y(\tau),\tau \,|\, \xi, s)  d\xi =
-\frac{\partial}{\partial \tau} \left( \frac{f^+(\tau)}{y(\tau)}\right) \Bigg\{ \\
&\frac{1}{2 y(\tau)} \left[ \theta_3\left( \phi_-(y(\tau), y(s), \tau), \omega_2(s, \tau) \right) + \theta_3\left( \phi_+(y(\tau), y(s), \tau), \omega_2(s, \tau) \right) \right] \nonumber \\
& + \frac{1}{2} \sum _{n=-\infty }^{\infty } \left[ \erf \left(\frac{(2 n +1) y(\tau ) + y(s)}{2 \sqrt{\tau-s }}\right)
- \erf \left(\frac{(2 n +1) y(\tau ) - y(s)}{2 \sqrt{\tau-s }}\right) \right] \Bigg\}.
\end{align}

The \eqref{finVolterra2} is nonlinear in $y(\tau)$ in contrast to a similar equation but w.r.t. $\Psi(\tau, y(\tau))$ which occurs when pricing barrier options, \citep{ItkinLiptonMuraveyBook}. Moreover, this is not a standard nonlinear Volterra integral equation, neither of the first nor of the second kind which by definition could be written in the form, \citep{polyanin2008handbook}
\begin{align} \label{VoltDef}
\xi(\tau) &= \int_0^\tau K(\tau, s, y(s)) ds, \\
y(\tau) &= \xi(\tau) + \int_0^\tau \calK(\tau, s, y(\tau), y(s)) ds, \nonumber
\end{align}
\noindent respectively, where $\xi(\tau)$ is some arbitrary function of $\tau$. In contrast, the integral Volterra equations  \eqref{finVolterra1} can be represented in a general form
\begin{equation} \label{nonLinVoltOU}
\xi(\tau, y(\tau)) = \int_0^\tau K(\tau, s, y(s)) ds.
\end{equation}
Below in this paper we discuss how to solve \eqref{finVolterra1} by using various methods also including those from machine learning (ML).

A short remark should be made that the dependencies like $a(t(\tau))$ can be computed by using some finite grid in $t \in [0,T]$ and then creating a map $t \to \tau \in [0, \tau(0)]$ according to the definition of $\tau(t)$ in \eqref{indV}. Since all these dependencies are analytic, this procedure doesn't bring any computational problem. Also, as $\tau = 0$ corresponds to $t=T$, it follows from \eqref{notation} that $y(0) = S_B(T) g(T) = K g(T)$ for the American Call option, while for the American Put this is $y(0) = g(T) \min(K, K r(T)/q(T))$, \citep{Kim:1990}

\subsection{The time-dependent Hull-White model} \label{tdHW}

This model was investigated in \citep{ItkinMuravey2020r} with a goal to construct semi-analytic prices of barrier options, in more detail see also \citep{ItkinLiptonMuraveyBook}. Therefore, in what follows we will borrow some results obtained in that paper. Historically, the model was introduced in \citep{Hull:1990a} to describe dynamics of the short interest rate $r_t$ as following the OU process with time-dependent coefficients
\begin{equation} \label{OU1-42}
d r_t = \kappa(t)[\theta(t) - r_t] dt + \sigma(t)dW_t, \qquad r_{t=0} = r.
\end{equation}
Here $\kappa(t) > 0$ is the constant speed of mean-reversion, $\theta(t)$ is the mean-reversion level. To address calibration to real market rate curves the model could be updated by using a deterministic shift $s(t)$, so $r(t) = s(t) + \bar{r}(t)$ where $\bar(t)$ solves \eqref{OU1-42}. This, however, can be easily done within our framework, as this is described in \citep{ItkinLiptonMuraveyBook}. In the commodities world this model is known as one-factor Schwartz's model, \citep{Schwartz1997}, for the logarithm of the spot price $S$.

\begin{figure}
\centering
\includegraphics[totalheight=2.7in]{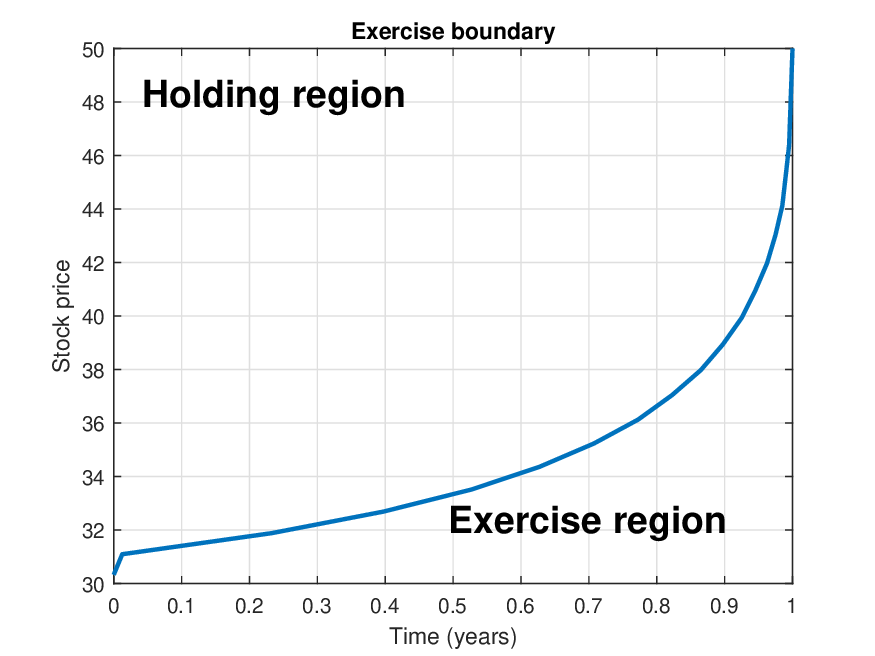}
\caption{Typical exercise boundary for the American Put option under the Black-Scholes model with $K = 50, r = 0.2, q = 0.1, \sigma = 0.5, T = 1$.}
\label{EBput}
\end{figure}

Since $r_t$ itself is not tradable, we consider a zero-coupon bond (ZCB) with the maturity $Q$ and the price $F(r,t,Q)$ as the underlying instrument and look at pricing of American options written on $F(r,t,Q)$, in particular, the American Put. A typical shape of the exercise boundary for the Put option is presented in Fig.~\ref{EBput}.

Using the same argument as in Section~\ref{tdOU}, let's assume that the exercise boundary $F_B(t)$ is known. Then, in the continuation (holding) region the pricing problem for the American Put $P(t,r)$ is equivalent to pricing Down-and-Out barrier option at the domain $\Omega: r \in [F(r_B(t), t), \infty) \times t \in [0,T]$, with $T \le Q$. It is known that in this region $F(r,t,Q)$ under a risk-neutral measure solves a linear parabolic partial differential equation (PDE), \citep{privault2012elementary, ItkinMuravey2020r}
\begin{equation} \label{PDE-42}
\fp{F}{t} + \dfrac{1}{2}\sigma^2(t) \sop{F}{r} + \kappa(t) [\theta(t) - r] \fp{F}{r} = r F.
\end{equation}
\noindent subject to the terminal
\begin{equation} \label{termZCB-42}
 F(r,Q,Q)  = 1,
\end{equation}
\noindent and boundary conditions
\begin{equation} \label{bc1-42}
 F(0,t,Q)  = g(t), \qquad  F(r,t,Q)\Big|_{r \to \infty} = 0,
\end{equation}
\noindent where $g(t)$ is some function of the time $t$, see, e.g., \citep{ZhangYang2017} and references therein among others. However,
since the Hull-White model belongs to the class of affine models, \citep{andersen2010interest}, the solution of \eqref{PDE-42} can be represented in the form
\begin{align} \label{affSol}
F(r,t,Q) &= A(t,Q) e^{ B(t,Q) r}, \\
B(t,Q) &= e^{\int_0^t \kappa (x) \, dx} \int_Q^t e^{-\int_0^x \kappa(q) \, dq} \, dx, \nonumber \\
A(t,Q) &= \exp\left[-\frac{1}{2} \int_Q^t  B(x,Q) \left( 2 \theta(x) \kappa(x) + B(x,Q) \sigma^2(x) \right) dx \right]. \nonumber
\end{align}
It can be seen that $B(t,Q) < 0$ if $t < Q$. Therefore,  $F(r,t,Q) \to 0$ when $r \to \infty$.

In turn, the American Put price $P(t,r)$ in the continuation region solves the PDE
\begin{equation} \label{PDEP}
\fp{P}{t} + \dfrac{1}{2}\sigma^2(t) \sop{P}{r} + \kappa(t) [\theta(t) - r] \fp{P}{r} = r P, \qquad (\tau , x) \in \mathbb{R}_+ \times [y(\tau), \infty),
\end{equation}
\noindent subject to the terminal condition
\begin{equation} \label{tchw}
P(T,r) = \left(K - F(r,T,Q)\right)^+,
\end{equation}
\noindent the boundary condition at the moving boundary
\begin{equation} \label{bc1hw}
P(t,r_B(t)) = K - F(r_B(t), t, Q) = K - A(t,Q) e^{ B(t,Q) r_B(t)},
\end{equation}
\noindent and the other boundary condition at the second boundary at $r \to \infty$. Since based on \eqref{affSol} in this limit the ZCB price tends to zero, this yields
\begin{equation} \label{bc2hw}
 P(t,r)\Big|_{r \to \infty}  = K.
\end{equation}

\subsubsection{Reduction to the heat equation}

By changing the dependent variable
\begin{equation} \label{chg1}
P(t,r) \to p(t,r) + K - F(r_B(t), t, Q),
\end{equation}
\noindent we obtain a similar problem but with a homogeneous boundary conditions at the moving boundary, inhomogeneous boundary condition at infinity and a source term, namely:
\begin{align} \label{PDEP1}
\fp{p}{t} &+ \dfrac{1}{2}\sigma^2(t) \sop{p}{r} + \kappa(t) [\theta(t) - r] \fp{p}{r} = r p + \bar{\lambda}(t,r), \\
\bar{\lambda}(t, r) &= r \left[ K - F(r_B(t), t, Q) \right] + \partial_t F(r_B(t), t, Q). \nonumber
\end{align}
\noindent This PDE has to be solved subject to the terminal condition
\begin{equation} \label{tchw2}
p(T,r) = \left(K - F(r, T, Q)\right)^+ - K + F(r_B(T), T, Q) =  0,
\end{equation}
\noindent which is due to the fact that in the continuation region for the American Put option $\left(K - A(T,Q)e^{B(T,Q) r}\right)^+ = 0$, see Fig.~\ref{EBput}, and also $A(T,Q) e^{B(T,Q) r_B(T)} = K$. Also, the boundary condition at the moving boundary becomes homogeneous
\begin{equation} \label{bc1hw2}
p(t, r_B(t)) = 0,
\end{equation}
\noindent and the other boundary condition at infinity - inhomogeneous
\begin{equation} \label{bc2hw2}
 p(t,r)\Big|_{r \to \infty}  = F(r_B(t), t, Q).
\end{equation}

This problem (with no source term $\lambda(t)$) has been already considered in \citep{ItkinMuravey2020r}. By transformations
\begin{equation} \label{transH}
p(t,r) = \exp[\alpha(t) r + \beta(t)] u(\tau, x), \qquad \tau = \phi(t), \qquad x = r \psi(t) + \xi(t),
\end{equation}
\noindent it can be reduced to the heat equation with a source term
\begin{align} \label{Heat2}
\fp{u}{\tau} &= \sop{u}{x} + \tilde{\lambda}(\tau, x), \\
\tilde{\lambda}(\tau, x) &= \frac{2 \bar{\lambda}(t(\tau), r)}{\sigma^2(t) \psi^2(t)}\exp\left[- \beta(t) - \frac{\alpha(t)}{\psi(t)}(x - \xi(t))\right], \quad t = t(\tau), \quad \ r = \frac{x - \xi(t)}{\psi(t)}, \nonumber
\end{align}
\noindent that should be solved subject to the initial and boundary conditions
\begin{align} \label{tc01}
u(0,x) &= 0, \qquad u(\tau, y(\tau)) = 0, \\
u(\tau, x)\Big|_{x \to \infty}  &= A(t,Q) \exp[\frac{B(t,Q) - \alpha(t)}{\psi(t)}(y(\tau) - \xi(t)) - \beta(t)] \equiv -g(\tau), \quad t = t(\tau). \nonumber
\end{align}
Here $y(\tau)$ is the exercise boundary expressed in new variables $(\tau, x)$, i.e. $y(\tau) = \xi(t(\tau)) + \psi(t) r_B(t(\tau))$.

The remaining notation in \eqref{Heat2} reads
\begin{alignat}{2} \label{coef}
\psi(t) &= \exp\left( \int_0^t \kappa(q) dq \right),  &\quad
\phi(t) &= \frac{1}{2} \int_t^T  \sigma^2(q) \psi^2(q) d q, \\
\alpha(t) &= \psi(t) \int_0^t \frac{1}{\psi(q)} dq, &\quad
\beta(t) &= - \frac{1}{2} \int_0^t \alpha(q) \left[2 \kappa(q) \theta(q) + \sigma^2(q) \alpha(q) \right] dq, \nonumber \\
\xi(t) &= -\int_0^t  \left[ \kappa(q) \theta(q) + \sigma^2(q) \alpha(q) \right] \psi(q) dq,
&\quad
\tau(t) &= \frac{1}{2} \int_t^T  \sigma^2(q) \psi^2(q) d q. \nonumber
\end{alignat}
Here $t(\tau)$ is the inverse map $t \rightarrow \tau$ which is found explicitly by solving the second equation in \eqref{coef} with respect to $t$.

\subsubsection{Nonlinear integral Volterra equation for $y(\tau)$} \label{voltHWPut}

In this section we present a GIT based approach to derive a Volterra integral equation for the exercise boundary $y(\tau)$.  For doing so, we start with a problem closely related to \eqref{Heat2} but with different boundary conditions. By slightly changing the notation: $u(\tau, x) \mapsto U(\tau, x)$, we get
\begin{align} \label{HeatPDE}
 \fp{U}{\tau} &=   \sop{U}{x}+ \lambda(\tau, x) \in \mathbb{R}_+ \times (y(\tau), +\infty) , \\
U(0, x) &= f(x), \qquad U(\tau, y(\tau)) = g(\tau),  \quad U(\tau, x)\Big|_{x \to +\infty} = 0. \nonumber
\end{align}
Here the functions $g(\tau)$, $f(x), \lambda(\tau,x)$ and $y(\tau)$ are defined to guarantee the existence and uniqueness of the problem \eqref{HeatPDE}. By the change of variables
\begin{equation} \nonumber
	V(\tau, x) = U(\tau, x) - g(\tau)
\end{equation}
\noindent we obtain the problem with absorbing boundary conditions (compare with \eqref{Heat2})
\begin{align} \label{HeatPDE_forV}
	 \sop{V}{x} &= \fp{V}{\tau} + g'(\tau) -  \lambda(\tau, x), \quad (\tau , x) \in \mathbb{R}_+ \times (y(\tau), +\infty) , \\
	V(\tau, y(\tau)) &= 0, \quad V(\tau, x)\Big|_{x \to +\infty} = -g(\tau), \nonumber  \\
	V(0, x) &= f(x) - g(0), \nonumber
\end{align}
By using the idea of GIT, let us consider a pair  of integral transforms $\bar v _\pm $
\begin{equation}  \label{GIT}
	\bar v _\pm  (\tau, \omega)= \int_{y(\tau)}^{+\infty} e^{\pm \iu \omega x } V(\tau, x) dx
\end{equation}
\noindent and apply them to \eqref{HeatPDE_forV}. We have
\begin{align*}
	\int_{y(\tau)}^{+\infty} e^{\pm \iu \omega x } \sop{V(\tau, x)}{x} dx &=  -\fp{V(\tau, x)}{x} \Big |_{x = y(\tau)} e^{\pm \iu \omega y(\tau) }
	\mp \iu \omega  	\int_{y(\tau)}^{+\infty} e^{\pm \iu \omega x } \fp{V(\tau, x)}{x} dx
	= -\Psi(\tau, y(\tau)) e^{\pm \iu \omega y(\tau)} - \omega^2 \bar v _{\pm}, \\
	\int_{y(\tau)}^{+\infty} e^{\pm \iu \omega x } \fp{V(\tau, x)}{\tau} dx &= \frac{d}{d \tau} \left[ 	\int_{y(\tau)}^{+\infty} e^{\pm \iu \omega x } V(\tau, x) dx\right] + e^{\pm \iu \omega x} V(\tau, y(\tau)) y'(\tau)
	= \frac{d \bar v _\pm}{d \tau}.
\end{align*}

Here the function $\Psi(\tau, y(\tau))$ is the gradient of the solution at the moving boundary $y(\tau)$
\begin{equation}
	\Psi(\tau, y(\tau)) =  \fp{V(\tau, x)}{x} \Big |_{x = y(\tau)}.
\end{equation}
Collecting all the terms together we obtain the ordinary differential equation (ODE) for $\bar v _\pm $
\begin{align} \label{vODE}
	\frac{d \bar v_\pm (\tau, \omega)}{ d\tau}  +\omega^2 \bar v _\pm &= -\Psi(\tau, y(\tau)) e^{\pm \iu \omega y(\tau)}
	- g'(\tau) 	\int_{y(\tau)}^{+\infty} e^{\pm \iu \omega x } dx + 	\int_{y(\tau)}^{+\infty} e^{\pm \iu \omega x } \lambda(\tau, x)dx   \\
	\bar v(0, \omega) &= \int_{y(0)}^{+\infty} e^{\pm \iu \omega x } (f(x)  - g(0)) dx. \nonumber
\end{align}
Here $y(0)$ is the value of the exercise boundary at maturity, which for an American Put option is, \citep{KaratzasShreeve1998}
\begin{equation}
y(0^+) =
\begin{cases}
K, & r(0) \ge q(0), \\
\frac{r(0)}{q(0)} K, & r(0) < q(0).
\end{cases}
\end{equation}

Solving \eqref{vODE} by using a standard technique yields the explicit formula for $\bar v (\tau, \omega)$
\begin{align} \label{barv_explicit}
	\bar v _\pm(\tau, \omega) &= e^{-\omega^2 \tau} \int_{y(0)}^{+\infty} e^{\pm \iu \omega x } (f(x)  - g(0)) dx  \\
	&- \int_0^\tau e^{-\omega^2 (\tau -s)}
	\left[
	\Psi(s, y(s)) e^{\pm \iu \omega y(s) } + g'(s) \int_{y(s)}^{+\infty} e^{\pm \iu \omega x } dx -
	  \int_{y(s)}^{+\infty} \lambda(s, x)dx
	\right] ds. \nonumber
\end{align}

To the best of our knowledge, each transform $\bar v _+$ and $\bar v _-$ is not explicitly invertible, however their linear combination is. Indeed,
\begin{align*}
	\bar v(\tau ,\omega) = \frac{\bar v_+ e^{-\iu \omega y(\tau)} - \bar v_- e^{+\iu \omega y(\tau)} }{2 \iu} = \frac{1}{\iu} \int_{y(\tau)} ^{+\infty}  V(\tau, x )
	\sinh \left[ \iu \omega  (x- y(\tau)) \right] dx  \\
	= \int_{y(\tau)}^{+\infty} V(\tau, x) \sin \left( \omega (x- y(\tau)) \right) dx
	=\int_{0}^{+\infty} V(\tau, x + y(\tau)) \sin \left( \omega x \right) dx
\end{align*}
\noindent which is a sine transform with a well-known inversion formula. On the other hand, the function $\bar v (\tau, \omega)$ can be represented as following
\begin{align*}
	\bar v (\tau, \omega) &= e^{-\omega^2 \tau} \int_{y(0)}^{+\infty} \sin \left[ \omega (x - y(\tau)) \right] (f(x)  - g(0)) dx  \\
	& - \int_0^\tau e^{-\omega^2 (\tau -s)}
	\left[
	\Psi(s, y(s)) \sin \left[ \omega (y(s) - y(\tau)) \right]  +  \int_{y(s)}^{+\infty} \left(g'(s) - \lambda(s, x) \right) \sin \left[ \omega (x - y(\tau)) \right]  dx
	\right] ds.
\end{align*}

Integrating by parts and replacing $x$ by $\xi$ yields
\begin{align*}
	\bar v (\tau, \omega) & = e^{-\omega^2 \tau} \int_{y(0)}^{+\infty} \sin \left[ \omega (\xi - y(\tau)) \right] (f(\xi)  - g(0)) d\xi
	- \int_0^\tau e^{-\omega^2 (\tau -s)} 	\Psi(s,y(s)) \sin \left[ \omega (y(s) - y(\tau)) \right]  ds \\
	&-g(\tau)  \int_{y(\tau)}^{+\infty} \sin \left[ \omega (\xi - y(\tau)) \right]  d\xi  + g(0) e^{-\omega^2 \tau} \int_{y(0)}^{+\infty} \sin \left[ \omega (\xi - y(\tau)) \right]  d\xi
	\\
	&+ \int_0^\tau  g(s) \frac{d}{ ds}
	\left[
	e^{-\omega^2 (\tau -s)} \int_{y(s)}^{+\infty}\sin \left(\omega (\xi - y(\tau)) \right) d\xi ds
	\right]
	\\
	&+  \int_0^\tau  e^{-\omega^2 (\tau -s)} \int_{y(s)}^{+\infty} \lambda(s, x)\sin \left(\omega (\xi - y(\tau)) \right) d\xi ds
	\\
	& = e^{-\omega^2 \tau} \int_{y(0)}^{+\infty} \sin \left[ \omega (\xi - y(\tau)) \right] f(\xi) d\xi
	- \int_0^\tau e^{-\omega^2 (\tau -s)} 	\Psi(s,y(s)) \sin \left[ \omega (y(s) - y(\tau)) \right]  ds \\
	&-g(\tau)  \int_{y(\tau)}^{+\infty} \sin \left[ \omega (x - y(\tau)) \right]  dx
	+ \int_0^\tau  g(s) \frac{d}{ ds}
	\left[
	e^{-\omega^2 (\tau -s)} \int_{y(0)}^{+\infty}\sin \left(\omega (\xi - y(\tau)) \right) d\xi ds
	\right]
	\\
		&+  \int_0^\tau  e^{-\omega^2 (\tau -s)} \int_{y(s)}^{+\infty} \lambda(s, \xi)\sin \left(\omega (\xi - y(\tau)) \right) d\xi ds
\end{align*}
Using the inversion formula for the sine transform
\begin{equation} \label{sine_transform_inversion}
	V(\tau, x) = \frac 2\pi \int_{0} ^{+\infty} \bar v(\tau, \omega) \sin \left[\omega (x - y(\tau))\right] d\omega .
\end{equation}
\noindent we obtain
\begin{align} \label{Vfinal}
	V(\tau, x) &= \frac 2\pi  \int_{0}^{+\infty} \sin\left(\omega(x - y(\tau)) \right)e^{-\omega^2 \tau} \int_{y(0)}^{+\infty} \sin \left[ \omega (\xi - y(\tau)) \right] f(\xi) d\xi d\omega \\
	&- \frac 2\pi  \int_{0}^{+\infty} \sin\left(\omega(x - y(\tau)) \right)  \int_0^\tau e^{-\omega^2 (\tau -s)} 	\Psi(s,y(s)) \sin \left[ \omega (y(s) - y(\tau)) \right]  ds d\omega \nonumber \\
	&- \frac 2\pi  \int_{0}^{+\infty} \sin\left(\omega(x - y(\tau)) \right) g(\tau)  \int_{y(\tau)}^{+\infty} \sin \left[ \omega (\xi - y(\tau)) \right]  d\xi  d\omega  \nonumber \\
	&+  \frac 2\pi  \int_{0}^{+\infty} \sin\left(\omega(x - y(\tau)) \right) \int_0^\tau  g(s) \frac{d}{ ds}
	\left[
	e^{-\omega^2 (\tau -s)} \int_{y(0)}^{+\infty}\sin \left(\omega (\xi - y(\tau)) \right) d\xi ds
	\right] d\omega \nonumber  \\
		&+ \frac 2\pi  \int_{0}^{+\infty} \sin\left(\omega(x - y(\tau)) \right) \int_0^\tau
		\lambda(s, \xi)
	e^{-\omega^2 (\tau -s)} \int_{y(0)}^{+\infty}\sin \left(\omega (\xi - y(\tau)) \right) d\xi ds
	d\omega. \nonumber
\end{align}
By changing the order of integration and applying  well-known identities, \citep{GR2007}
\begin{equation*}
	\int_0^{+\infty} e^{-\beta^2 x} \sin ( a x ) \sin( b x ) dx = \frac{1}{4} \sqrt{\frac \pi \beta} \left( e^{-\frac{(a-b)^2}{4\beta}} -  e^{-\frac{(a+b)^2}{4\beta}}\right)
\end{equation*}
\begin{equation*}
	\frac 2\pi  \int_{0}^{+\infty} \sin\left(\omega(x - y(\tau)) \right)  \int_{y(\tau)}^{+\infty} \sin \left[ \omega (\xi - y(\tau)) \right]  d\xi  d\omega
	\equiv  1.
\end{equation*}
\noindent we arrive at the following formula
\begin{align*}
	V(\tau, x) = &- g(\tau) + \frac{1}{2\sqrt{\pi \tau}} \int_{y(0)}^{+\infty} f(\xi ) \left[e^{-\frac{( \xi - x)^2}{4\tau}} -  e^{-\frac{(\xi + x -2 y(\tau))^2}{4\tau}} \right] d\xi  \\
	&-\int_0^\tau \frac{\Psi(s,y(s))}{2\sqrt{\pi (\tau - s)}} \left[e^{-\frac{(x - y(s))^2}{4(\tau-s)}} -  e^{-\frac{(x -2 y(\tau) + y(s))^2}{4(\tau-s)}} \right] ds\\
	&+ \int_0^\tau g(s) \left[ \frac{d}{ ds} \left(\frac{1}{2\sqrt{\pi (\tau - s)}}
	\int_{y(s)}^{+\infty}\left[e^{-\frac{(x - \xi)^2}{4(\tau-s)}} -  e^{-\frac{(x -2 y(\tau) + \xi))^2}{4(\tau-s)}} \right]d\xi \right)\right] ds \\
 	&+\int_0^\tau \int_{y(s)}^{\infty}\frac{\lambda(s, \xi)}{2\sqrt{\pi (\tau - s)}} \left[e^{-\frac{(\xi - y(s))^2}{4(\tau-s)}} -  e^{-\frac{(\xi -2 y(\tau) + y(s))^2}{4(\tau-s)}} \right]  d\xi ds
\end{align*}

Substituting back $U = V + g(\tau)$ and doing some algebra yields
\begin{align*}
	U(\tau, x) &= \frac{1}{2\sqrt{\pi \tau}} \int_{y(0)}^{+\infty} f(\xi ) \left[e^{-\frac{( \xi - x)^2}{4\tau}} -  e^{-\frac{(\xi + x -2 y(\tau))^2}{4\tau}} \right] d\xi  \\
	&-\int_0^\tau \frac{\Psi(s,y(s))}{2\sqrt{\pi (\tau - s)}} \left[e^{-\frac{(x - y(s))^2}{4(\tau-s)}} -  e^{-\frac{(x -2 y(\tau) + y(s))^2}{4(\tau-s)}} \right] ds\\
	&+ \int_0^\tau g(s) \left[ \frac{d}{ ds} \left(\frac{1}{2\sqrt{\pi (\tau - s)}}
	\int_{y(s)}^{+\infty}\left[e^{-\frac{(x - \xi)^2}{4(\tau-s)}} -  e^{-\frac{(x -2 y(\tau) + \xi))^2}{4(\tau-s)}} \right]d\xi \right)\right] ds \\
	&+\int_0^\tau \int_{y(s)}^{\infty}\frac{\lambda(s, \xi)}{2\sqrt{\pi (\tau - s)}} \left[e^{-\frac{(\xi - y(s))^2}{4(\tau-s)}} -  e^{-\frac{(\xi -2 y(\tau) + y(s))^2}{4(\tau-s)}} \right]  d\xi ds
\end{align*}

Again, integrating by parts in the last integral, we finally obtain
\begin{align} \label{U_final}
	U(\tau, x) &= \frac{1}{2\sqrt{\pi \tau}} \int_{y(0)}^{+\infty} f(\xi ) \left[e^{-\frac{( \xi - x)^2}{4\tau}} -  e^{-\frac{(\xi + x -2 y(\tau))^2}{4\tau}} \right] d\xi  \\
	&-\int_0^\tau \frac{\Psi(s,y(s)) + y'(s) g(s)}{2\sqrt{\pi (\tau - s)}} \left[e^{-\frac{(x - y(s))^2}{4(\tau-s)}} -  e^{-\frac{(x -2 y(\tau) + y(s))^2}{4(\tau-s)}} \right] ds  \nonumber \\
	&+ \int_0^\tau \frac{g(s)}{4\sqrt{\pi(\tau - s)^3}} \left[
	(x- y(s)) e^{-\frac{(x-y(s))^2}{4(\tau -s)}} +
	(y(s)+ x - 2 y(\tau)) e^{-\frac{(x+y(s) - 2 y(\tau))^2}{4(\tau -s)}} \right] ds \nonumber \\
		&+\int_0^\tau \int_{y(s)}^{\infty}\frac{\lambda(s, \xi)}{2\sqrt{\pi (\tau - s)}} \left[e^{-\frac{(\xi - x)^2}{4(\tau-s)}} -  e^{-\frac{(\xi -2 y(\tau) + x)^2}{4(\tau-s)}} \right]  d\xi ds. \nonumber
\end{align}

Thus, \eqref{U_final} gives a closed form solution of the problem \eqref{HeatPDE}. It is worth noting that the last term in \eqref{Heat2} can be considered as an application of the Duhamel's principle to the problem \eqref{Heat2}. However, we didn't find in the literature any reference to using the Duhamel's formula for problems with moving boundaries.

To apply this result to our problem of pricing the American Put option in \eqref{Heat2}, we need to set
\begin{align} \label{AmerPutCond}
u(\tau, x) &= V(\tau,x) = U(\tau,x)  - g(\tau), \qquad f(x) = g(0), \qquad
\lambda(\tau, x) = \tilde{\lambda}(\tau,x) - g'(\tau),
\end{align}
\noindent where $g(\tau)$ is defined in \eqref{tc01}. Accordingly, $f'(x) = 0$ and the first term in \eqref{U_final} vanishes. Also, we have from \eqref{PDEP1}, \eqref{Heat2}, \eqref{tc01}
\begin{align*}
\lambda(\tau,x) &= \frac{2 r e^{-r \alpha(t) - \beta(t) }}{\sigma(t)^2 \psi(t)^2}  \Bigg\{
K - F(r_B(t), t, Q) + e^{B(t,Q) r_B(t)} \left[ f_1(\tau,y(\tau)) e^{\alpha(t) (x - y(t))/\psi(t)} + f_2(\tau, y(\tau)) \right] \Bigg\}, \\
r &= \frac{x - \xi(t)}{\psi(t)}, \quad t = t(\tau), \quad r_B(t) = \frac{y(\tau) - \xi(t(\tau))}{\psi(t(\tau))}.
\end{align*}
\noindent where expressions for $f_1(\tau, y(\tau)), f_2(\tau, y(\tau))$ are simple but bulky, so we are omitting them here. Therefore, the last integral in \eqref{U_final} simplifies since, e.g.,
\begin{align*}
\int_{y(s)}^{\infty}\frac{1}{2\sqrt{\pi (\tau - s)}} e^{-\frac{(\xi - x)^2}{4(\tau-s)}} d\xi &= \frac{1}{2} \left[1 + \erf\left(\frac{x- y(s)}{2 \sqrt{\tau -s}}\right) \right],
\end{align*}
\noindent etc., (i.e., all the integrals of a product of $\lambda(s,\xi)$ and the Gaussian kernel are expressed in closed form via $\erf, \erfc, \erfi$ functions).

Similarly to \eqref{finVolterra1}, the exercise boundary $y(\tau)$ solves a nonlinear Volterra integral equation, that can be obtained from \eqref{U_final} by differentiating both parts on $x$ and setting $x = y(\tau)$. This equation also connects the gradient at the moving boundary $\Psi(\tau, , y(\tau))$ and the exercise boundary $y(\tau)$. However, due to existing singularities in the integrals in \eqref{U_final} this procedure requires a special approach which is described in Appendix~\ref{app2}. It is shown there that the corresponding (regularized) Volterra integral equation reads
\begin{align}  \label{finVoltalt}
	\Psi(\tau, y(\tau)) &= \int_{y(0)}^\infty f'(\xi ) \frac{e^{-\frac{(\xi - y(\tau))^2}{4\tau}}}{\sqrt{\pi \tau}}  d\xi +
	\int_0^\tau \Psi(s, y(s)) \frac{ (y(\tau) - y(s))  e^{-\frac{(y(\tau) - y(s))^2}{4(\tau - s)}}  }{2\sqrt{\pi (\tau - s)^3}} ds
	-\int_0^\tau g'(s)  \frac{e^{-\frac{(y(\tau) - y(s))^2}{4(\tau - s)}} }{\sqrt {\pi (\tau - s) }}   ds  \nonumber \\
	&+\int_0^\tau \int_{y(s)}^{\infty} \lambda(s, \xi) \frac{\xi - y(\tau)}{2\sqrt{\pi (\tau - s)^3}} e^{-\frac{(\xi - y(\tau))^2}{4(\tau-s)}} d\xi ds.
\end{align}
It can be checked that this equation does not contain singularities anymore. Since for the American option the function $\Psi(\tau, y(\tau))$ is known,  \eqref{finVoltalt} is another form of the non-linear Volterra equation to be solved in order to compute the exercise boundary $y(\tau)$ for the American Put option in the time-dependent Hull-White model.

Again, as applied to pricing of the American Put option in \eqref{Heat2}, based on \eqref{AmerPutCond} we have the first integral in \eqref{finVoltalt} vanishing, and the last integral on $\xi$ can be taken in closed form, while for $\Psi(\tau, y(\tau)$ we have from \eqref{bc1hw}, \eqref{transH}, \eqref{tc01}
\begin{align} \label{PsiHW}
\Psi(\tau, y(\tau) &= \fp{V(\tau, x)}{x} \Big |_{x = y(\tau)} = \fp{u(\tau, x)}{x} \Big |_{x = y(\tau)}
= - A(t,Q) B(t,Q) e^{[B(t,Q) - \alpha(t)] r_B(\tau) - \beta(t)} = B(\tau, Q) g(\tau), \nonumber \\
t &= t(\tau), \quad r_B(t) = \frac{y(\tau) - \xi(\tau)}{\psi(\tau)}.
\end{align}

To conclude, given a) some time-dependent model for the underlying (e.g., the stock price), b) the corresponding pricing PDE for the American option written on this underlying which is valid in the continuation region, c) this PDE by a series of transformations can be reduced to the heat equation with moving boundaries, it follows that the exercise boundary solves
an integral non-linear Volterra equation that can be derived by using the method described in this section.

\subsection{The time-dependent Verhulst model} \label{secVerhulst}

A slightly more sophisticated example of the proposed technique is pricing American options under the time-dependent Verhulst model. As applied to finance it was proposed in \citep{ItkinLiptonMuravey2020} based on the observation that some popular short-term interest models have to be revisited and modified to better reflect market conditions during COVID-19 period. The model is a modification of the popular Black-Karasinski (BK) model, which is widely used by practitioners for modeling interest rates, credit, and commodities. This modification gives rise to the stochastic Verhulst model, which is well-known in the population dynamics and epidemiology as a logistic model.

The main idea behind the model is that since the BK model doesn't support fat tails at the lower end, besides not being analytically tractable, we introduce its modified version of the form
\begin{align} \label{BK}
d z_t &= \kappa(t)[\bar{\theta}(t) - e^{z_t}] dt + \sigma(t)dW_t, \\
r_t  &=  s(t) + R e^{z_t}, \qquad R = r_0 - s(0), \qquad (t,z_t) \in [0,T]\times(-\infty,\infty). \nonumber
\end{align}
In other words, we modify the dynamics of the stochastic variable $z_t$ in \eqref{BK} in the mean-reversion term by replacing $z_t$ with $e^{z_t}$.

In \eqref{BK} $R$ is some constant  with the same dimensionality as $r_t$, e.g., it can be $R = r(0) - s(0)$, T is the maturity. This model is similar to the Hull-White model, but preserves positivity of $r_t$ by exponentiating the OU random variable $z_t$. Because of that, usually practitioners add a deterministic function (shift) $s(t)$ to the definition of $r_t$ to address possible negative rates and be more flexible when calibrating the term-structure of the interest rates.

It can be seen, that at small $t$ $|z_t| \ll 1$, and so choosing $\bar{\theta}(t) = 1 + \theta(t)$ replicates the BK model in the linear approximation on $z_t$. Similarly, the choice $\bar{\theta}(t) = e^{\theta(t)}$ replicates the BK model at $z_t$ close the mean-reversion level $\theta(t)$. Thus, the Verhulst model acquires the properties of the BK model while is a bit more tractable, again in more detail see \citep{ItkinLiptonMuravey2020, ItkinLiptonMuraveyBook}.

By a change of variables \eqref{BK} transforms to the stochastic Verhulst or stochastic logistic model, which are well-known in the population dynamics and epidemiology; see, e.g., \citep{Verhulst, bacaer2011, Logistic2015} and references therein. In the past, several authors attempted to use this model in finance; see, e.g., \citep{Chen2010, Londono2015, Halperin2018}. In our case, the stochastic Verhulst equation has the form
\begin{align} \label{verhulst}
d \br_t &= k(t) \br_t [\tth(t) - \br_t] dt + \sigma(t) \br_t dW_t, \qquad
\br_t = \frac{r_t - s(t)}{r_0 - s(0)} = e^{z_t}.
\end{align}

Accordingly, by \Ito lemma and the Feynman–Kac formula any contingent claim written on the $r_t$ as the underlying (for instance, the price $F(\br,t,Q)$ of a ZCB with maturity $T$) solves the following PDE
\begin{align} \label{PDEBK}
0 &= \fp{F}{t} + \dfrac{1}{2}\sigma^2(t) \br^2 \sop{F}{\br} + \kappa(t) \br [\tth(t) - \br] \fp{F}{\br} - (s(t) + R \br)F,
\end{align}
which should be solved subject to the terminal and boundary conditions \footnote{
Since $\br \in (0,\infty)$, i.e., the boundary $\br = 0$ is not attainable, by Fichera's theory \eqref{PDEBK} doesn't need the boundary condition at the left boundary $\br \to 0$, i.e. the PDE itself with substituted $\br = 0$ serves as the boundary condition. Otherwise, the boundary condition at this point should be set as
$F(\bar{r},t,Q)|_{\bar{r} \to 0} \to \exp[\int_Q^t s(k) dk]$, \citep{ItkinLiptonMuravey2020}.}
\begin{equation} \label{termZCB2}
 F(\br, Q, Q)  = 1, \qquad F(\br, t, Q)\Big|_{\br \to \infty} = 0.
\end{equation}

In general, $F(\br, t, Q)$ can be found e.g., by solving an integral Volterra equation derived in \citep{ItkinLiptonMuravey2020}. However, in some cases a closed form expression can also be found. For instance, when $\kappa(t) = \kappa = const$, it was obtained in \citep{ItkinLiptonMuravey2020} and reads
\begin{align} \label{u_explicit_f4_new}
F(\tau,Q, w) & e^{C_a(C_a-1)\tau - \int_T^{t(\tau)} s(k) dk} = 1 +e^{w/2} w^{-C_\alpha} \left[ \calI(\tau,w)
+ \frac{1}{\pi^2 \Gamma (R/\kappa)} \calJ(\tau,w) \right], \\
\calJ(\tau,w) &=  \int_0^\infty \omega \left[e^{-(\omega^2  +1/4) \tau} - 1 \right] \sinh\left(2 \pi \omega\right) \Gamma_{C_\gamma}^-(\omega) \Gamma_{C_\gamma}^+(\omega) \Gamma_{1-C_\alpha}^-(\omega) \Gamma_{1-C_\alpha}^+(\omega) W_{C_\gamma, \iu \omega}\left(w\right) \, d \omega, \nonumber \\
\calI(\tau,w) &= \sum_{k = 0}^K \frac{e^{\left(\mu_k^2 - \frac{1}{4}\right) \tau} }{k!}
\frac{2\mu_k }{\Gamma\left( \frac 32 -C_\alpha + \mu_k\right)} \frac{\Gamma_{C_\gamma}^+\left(\iu \mu_k\right) \Gamma_{C_\gamma}^-\left(\iu \mu_k\right)}{ \Gamma(C_\alpha - C_\gamma)} W_{C_\gamma, \mu_k}\left(w\right).
\nonumber
\end{align}
Here $Q=Q(\tau)$, $W_{\kappa, \mu}(w), \, \Gamma(x)$ are the Whittaker and Gamma functions, \citep{as64}, $C_\alpha = const$ - a new parameter of the model that can be found by calibration of all the model parameters to market data
\footnote{Thus, in this version of the model, we have three calibration parameters: two of them - $\kappa$ and $C_\alpha$ are constants, and the normal volatility $\sigma(t)$ is time-dependent. In other words, this enables capturing the volatility term-structure of the market which seems to be the most important property, while assuming a constant mean reversion speed is not too restrictive. The time-dependence of the mean reversion level, however, is fully defined by $\sigma(t)$  and is corrected by another calibrated constant $C_\alpha$. So this seems to be a weak side of the model.}, and the following change of variables has been made
\begin{align}
w &= \frac{2\kappa \phi(t) }{\sigma^2(0)} \br,  \quad \tau = \frac{1}{2}\int_t^Q \sigma^2(k) \, dk,
\quad  \phi(t) = e^{\int_0^t\left[C_\alpha \sigma^2(k) - \kappa \tth(k)\right] dk},
\quad C_\gamma =  \frac{\gamma(t)}{\kappa}, \quad \mu_k = \frac 12 - k - C_\alpha.
\end{align}
Also, in \eqref{u_explicit_f4_new} $K$ denotes the number of poles of the Gamma function which depends on the value of $C_\alpha$ : if $C_\alpha \geq 1/2$, the integrand function has no poles, if $C_\alpha <1/2$ the number of the poles is $K = 1 + \left\lfloor\frac 12 - C_\alpha \right\rfloor$, where $\lfloor x \rfloor$ is the floor of $x$.

\subsubsection{Nonlinear integral Volterra equation for $y(\tau)$} \label{voltVerhPut}

Similar to Section~\ref{tdHW}, we consider an American option written on a zero-coupon bond $F(\br,t,Q)$. In the continuation (holding) region the pricing problem for the American Put $P(t,r)$ reads
\begin{align} \label{PDE-Ver}
\fp{P}{t} &+ \dfrac{1}{2}\sigma^2(t) \br^2 \sop{P}{\br} + \kappa(t) \br [\tth(t) - \br] \fp{P}{\br}  =
 (s(t) + R \br)P, \\
 P(T,\br) &= \left(K - F(\br,T,Q)\right)^+ = 0, \qquad
 P(t,\br_B(t)) = K - F(\br_B(t), t, Q), \qquad P(t,\br)\Big|_{\br \to \infty}  = K. \nonumber
 \end{align}

 Similar to the change of the dependent variable in \eqref{chg1} we do
 \begin{equation} \label{chg11}
P(t,r) \to p(t,r) + K - F(\br_B(t), t, Q),
\end{equation}
\noindent to obtain from \eqref{PDE-Ver}
\begin{align} \label{PDE-Ver1}
\fp{p}{t} &+ \dfrac{1}{2}\sigma^2(t) \br^2 \sop{p}{\br} + \kappa(t) \br [\tth(t) - \br] \fp{p}{\br}  =
 [s(t) + R \br] p + \bar{\lambda}(t,\br), \\
\bar{\lambda}(t, \br) &= [s(t) + R \br] \left[ K - F(\br_B(t), t, Q) \right] + \partial_t F(\br_B(t), t, Q), \nonumber \\
p(T,\br) &= 0, \qquad  p(t,\br_B(t)) = 0, \qquad p(t,\br)\Big|_{\br \to \infty}  = F(\br_B(t), t, Q). \nonumber
 \end{align}

 Using another transformation proposed in \citep{ItkinLiptonMuravey2020}
\begin{align}
p(t, \br) &= e^{\frac{\kappa(t)}{\sigma^2(t)} \br + \int_T^t s(k) dk} u(\tau, y), \quad
x = \log \br + \int_0^t \left[ \frac{\sigma^2(k)}{2} - \kappa(k) \tth(k) \right] dk, \quad \tau = \frac{1}{2}\int_t^T \sigma^2(k) dk,
\end{align}
\noindent this problem can be reduced to a simpler form
\begin{align} \label{PDE_ZCB_u}
\fp{u}{\tau} &=  \sop{u}{x} + \left[ \alpha(\tau) e^{x} + \beta(\tau) e^{2x}\right] u + \lambda(\tau,x), \\
\alpha(\tau) &= \frac{2 \gamma(t(\tau))}{\sigma^2(t(\tau))} e^{-\int_0^t \left[ \frac{\sigma^2(k)}{2} - \kappa(k) \tth(k) \right] dk}, \qquad \beta(\tau) = - \frac{\kappa^2(t(\tau))}{\sigma^4(t(\tau))} e^{-\int_0^t \left[\sigma^2(k) - 2\kappa(k) \tth(k) \right] dk}, \nonumber \\
\gamma(t) &= - R + \frac{\kappa^2(t) \tth(t)}{\sigma^2(t) } + \frac{d}{dt}\left(\frac{\kappa(t)}{\sigma^2(t)}\right),
\qquad \lambda(\tau,x) = \frac{2}{\sigma^2(t(\tau))}\bar{\lambda}(t(\tau), \br(x)), \nonumber \\
u(0,\br) &= 0, \qquad  u(\tau,y(\tau)) = 0, \qquad u(\tau,x)\Big|_{x \to \infty}  =
e^{-\frac{\kappa(t)}{\sigma^2(t)} \br(x) + \int_0^\tau s(k) dk} F(y(\tau),\tau, Q(\tau)). \nonumber
\end{align}

Let us compare two problems in \eqref{PDE_ZCB_u} and \eqref{HeatPDE_forV}. They differ only by the existence of an extra source term $\left[ \alpha(\tau) e^{x} + \beta(\tau) e^{2x}\right] u$ in the RHS of \eqref{HeatPDE_forV}. However, it is known that the Duhamel's principle still can be applied in this case as this is discussed in \citep{ItkinLiptonMuravey2020}. Therefore, we can use the solution of \eqref{PDE_ZCB_u} in \eqref{U_final} which will now have an extra term in the RHS
\begin{align} \label{U_final1}
	U(\tau, x) &= \frac{1}{2\sqrt{\pi \tau}} \int_{y(0)}^{+\infty} f(\xi ) \left[e^{-\frac{( \xi - x)^2}{4\tau}} -  e^{-\frac{(\xi + x -2 y(\tau))^2}{4\tau}} \right] d\xi  \\
	&-\int_0^\tau \frac{\Psi(s,y(s)) + y'(s) g(s)}{2\sqrt{\pi (\tau - s)}} \left[e^{-\frac{(x - y(s))^2}{4(\tau-s)}} -  e^{-\frac{(x -2 y(\tau) + y(s))^2}{4(\tau-s)}} \right] ds  \nonumber \\
	&+ \int_0^\tau \frac{g(s)}{4\sqrt{\pi(\tau - s)^3}} \left[
	(x- y(s)) e^{-\frac{(x-y(s))^2}{4(\tau -s)}} +
	(y(s)+ x - 2 y(\tau)) e^{-\frac{(x+y(s) - 2 y(\tau))^2}{4(\tau -s)}} \right] ds \nonumber \\
		&+\int_0^\tau \int_{y(s)}^{\infty}\frac{1}{2\sqrt{\pi (\tau - s)}} \left[e^{-\frac{(\xi - x)^2}{4(\tau-s)}} -  e^{-\frac{(\xi -2 y(\tau) + x)^2}{4(\tau-s)}} \right]  \left\{ \lambda(s, \xi) + \left[\alpha(s) e^{\xi} + \beta(s) e^{2\xi}\right] U(s,\xi) \right\} d\xi ds. \nonumber
\end{align}
Therefore, in this case \eqref{U_final1} is not a closed form solution anymore, but rather an integral equation for $U(\tau, x)$.

In a similar way a non-linear integral Volterra equation for the exercise boundary can be obtained from \eqref{finVoltalt}, which now also contains an extra term
\begin{align}  \label{finVoltalt2}
	\Psi(\tau, y(\tau)) &= \int_{y(0)}^\infty f'(\xi ) \frac{e^{-\frac{(\xi - y(\tau))^2}{4\tau}}}{\sqrt{\pi \tau}}  d\xi +
	\int_0^\tau \Psi(s, y(s)) \frac{ (y(\tau) - y(s))  e^{-\frac{(y(\tau) - y(s))^2}{4(\tau - s)}}  }{2\sqrt{\pi (\tau - s)^3}} ds - \int_0^\tau g'(s)  \frac{e^{-\frac{(y(\tau) - y(s))^2}{4(\tau - s)}} }{\sqrt {\pi (\tau - s) }}   ds  \nonumber \\
&+\int_0^\tau \int_{y(s)}^{\infty} \left\{ \lambda(s, \xi) + \left[\alpha(s) e^{\xi} + \beta(s) e^{2\xi}\right] U(s,\xi) \right\} \frac{\xi - y(\tau)}{2\sqrt{\pi (\tau - s)^3}} e^{-\frac{(\xi - y(\tau))^2}{4(\tau-s)}} d\xi ds.
\end{align}
Since the gradient of the solution $\Psi(\tau, y(\tau))$ is known in closed form, the equations \eqref{U_final1} and \eqref{finVoltalt2} form a system of two non-linear equations. By solving it, the American Put price in the continuation region and the exercise boundary can be found. Numerical solution can be constructed along the lines discussed below in Sections~\ref{interCom},~\ref{ML}, see also \citep{ItkinLiptonMuravey2020}. However, since \eqref{U_final1} is a two-dimensional equation, this method is less efficient compared to the other models considered in this paper.

\section{Some intermediate comments} \label{interCom}

Two examples presented in Sections~\ref{tdOU},~\ref{tdHW} reveal some general steps necessary to apply the GIT approach originally developed for pricing barrier options in the time-dependent models, to pricing American options for the same models. Due to generality of these steps, in this section we concentrate reader's attention at them also focusing on some important details.

The first comment is about a map between the barrier and American options. It was seen in Section~\ref{tdOU} that pricing American Call options due to the corresponding boundary conditions can be transformed to pricing an Up-And-Out barrier option in the continuation region defined at the domain $\Omega: S \in [0, S_B(t)] \times t \in [0,T]$, see Fig.~\ref{EB}. In other words, the spatial domain for this problem has a moving (time-dependent) upper boundary. As the GIT method has been developed for solving this kind of problems in a semi-analytical manner, due to the above-mentioned similarity it can be naturally applied for pricing American options as well.

For the American Put option doing in a similar manner as in Section~\ref{tdHW} one can reduce a pricing problem for the American option to that one for the Down-And-Out barrier option defined at the domain $\Omega: S \in [S_B(t), \infty] \times t \in [0,T]$, see Fig.~\ref{EBput}. Again, the GIT method can be utilized to construct a semi-analytic solution of this problem as this was shown in \citep{ItkinLiptonMuraveyBook} and various papers of the authors referenced therein.

The second point to be emphasized has been already discussed in Introduction and is about the boundary condition at the exercise boundary $S_B(t)$. For barrier options this is usually either zero (so the contract is terminated at hit of the barrier), or some constant rebate paid either at hit or at option's maturity. In contrast, for the American option the payoff at the exercise boundary is known and is either $C_A(S_B(t),t)) = S_B(t) - K$ for the Call, and $P_A(S_B(t),t)) = K - S_B(t)$ for the Put. Accordingly, the option Delta is also known and is either 1 for the Call, or -1 for the Put.

As has been demonstrated in Section~\ref{tdHW}, by a simple change of variables this problem can be reduced to that with homogeneous boundary conditions (to exactly mimic the settings considered in most of the problems in \citep{ItkinLiptonMuraveyBook})\footnote{Inhomogeneous boundary conditions have been also considered in \citep{ItkinMuraveyDBFMF}.}. The cost for doing so is that the pricing PDE now becomes inhomogeneous and acquires a source term, see \eqref{Heat2}. However, given the solution of the {\it homogeneous} PDE, the full solution of the corresponding  {\it inhomogeneous} PDE can be found either directly by using the GIT method, or by using a generalized Duhamel's principle, like this was done in Section~\ref{voltHWPut}.

It should be emphasized that, despite so far we described our method only for the time-dependent OU and Hull-White  models, any other model where the pricing PDE can be reduced to the heat equation with a source term can be treated in the same way. For instance, pricing Amercian Call and Put options in the time-dependent Black-Scholes model can be done by first, reducing the pricing PDE in the continuation region to the form of \eqref{HeatPDE}, and then using the solution of this equation obtained in \eqref{U_final}. All these models can be reversely obtained from the heat equation  by using Lie group analysis, see \citep{Olver1993, Gazizov1998} among others.

The third point also has been already briefly discussed in Section~\ref{voltOU} and is about a Volterra integral equation which the unknown exercise boundary $S_B(t)$ solves. In contrast to the barrier options where a similar Volterra equation is derived for the option Delta at the moving boundary and is linear in Delta, here Volterra equations for $S_B(t)$ are non-linear and also of a special form given in \eqref{nonLinVoltOU}. To make it transparent we can slightly re-write \eqref{nonLinVoltOU} in a form
\begin{equation} \label{nonLinVoltOU1}
\xi(\tau, y(\tau)) - \xi(\tau, y(0)) = \int_0^\tau K(\tau, s, y(s)) ds - \xi(\tau, y(0)),
\end{equation}
\noindent and assuming the function $\xi(\tau, x)$ has all derivatives in $x$ to be finite, obtain from Tailor series in $x = y(\tau)$
\begin{align} \label{Tailor}
\int_0^\tau K(\tau, s, y(s)) ds - \xi(\tau, y(0)) = \xi'_x(\tau, x)\Big|_{x=y(0)} (y(\tau) - y(0)) + \text{extra terms}, \end{align}
\noindent where, e.g., for the time-dependent OU model considered in Section~\ref{tdOU}: $y(0) = K g(T)$. With no extra terms in the RHS, \eqref{Tailor} has a standard form of the non-linear Volterra equation of the second kind, but extra terms add more complexity to this equation.

It is worth noticing that \eqref{nonLinVoltOU} can also be re-written in the standard form
\begin{align} \label{stVolt2}
y(\tau) = \xi_1(\tau) + \int_0^\tau K_1(\tau, s, y(s), y'(s)) ds,
\end{align}
\noindent with $K_1$ being the new kernel which, however, now also depends on $y'(s)$. In our opinion, from numerical point of view this form is less preferable since computation of $y'(s)$ in the kernel decreases an accuracy of any corresponding method.

The last portion of comments in this Section is devoted to traditional methods of solving non-linear integral equations. For instance, a straightforward approach of solving \eqref{finVolterra2} written in the form of \eqref{nonLinVoltOU} could be as follows. Let us represent the unknown function $y(s)$ on a discrete grid in time $\Omega: s \in [0, \tau]$ with $M+1$ nodes equally distributed at $[0, \tau]$ with the step $\delta = \tau/M)$. Thus, instead of a continuous variable $y(s)$  we obtain a discrete vector $y_i, \ i=0,\ldots,M$, such that $y_0 = y(0) = K g(T), \ y_{M} = y(\tau)$. The integral kernel in the RHS of \eqref{finVolterra2} can be approximated on this grid by using  some quadratures, e.g., the trapezoidal rule (to provide the final solution with the accuracy $O(\delta^2)$). Then, for each value $\tau = j \delta, j \in [0,\ldots, M]$ at the grid $\Omega$ the \eqref{nonLinVoltOU} we obtain
\begin{align} \label{seqNLalg}
0 = \delta \sum_{i=0}^{j} K(j \delta, i \delta, y_i) \left(1 - \frac{1}{2}\Ind_{i=0} \Ind_{i=j} \right) - \xi(j \delta, y_j),
\end{align}
\noindent where $\Ind_x$ is the indicator function. This is a system on nonlinear algebraic equations w.r.t. the vector of unknown values $y_1,\ldots,y_j$. Since $y_0$ is already known, it can be solved sequentially, first for $j=1$, then for $j=2$ as the value of $y_1$ is already known from the previous step, etc. For solving non-linear algebraic equations various standard method are available, see \citep{Rheinboldt1998, benton2018nonlinear} among others. Various programming languages (Python, Matlab, etc.) contain standard packages for solving nonlinear systems of equations.

To achieve a better accuracy, high order quadratures can be used for the kernel's approximation, e.g., the Simpson rule which allows finding the solution with the accuracy $O(\delta^4)$. In this case again  \eqref{stVolt2} can be solved sequentially, but only starting from $j > 2$. For $j=1,2$ the equations in \eqref{stVolt2} have to be solved together as a system of two nonlinear equations.

To illustrate this approach, we use an example from \citep{CarrItkin2020jd} where without any loss of generality parameters of the time-dependent OU model are chosen in the following way
\begin{equation} \label{ex}
r(t) = r_0 e^{- r_k t}, \qquad q(t) = q_0, \qquad \sigma(t) = \sigma_0 e^{-\sigma_k t}.
\end{equation}
Here $r_0, q_0, \sigma_0, r_k, \sigma_k$ are constants. With this model \eqref{Ric} can be solved analytically to yield
\begin{equation}
w(t) = q_0 - r_0 e^{-r_k t} - r_k e^{-2 \frac{r_0}{r_k} e^{- r_k t} } / E_{1 + 2 (q_0-\sigma_k)/r_k}
\left( -2 \frac{r_0}{r_k} e^{- r_k t} \right),
\end{equation}
\noindent where $E_n(z)$ is the exponential integral function $E_n(z) = \int_1^\infty e^{- z t}/t^n d t$, \citep{as64}.

We choose parameters of the model as they are presented in Table~\ref{tab1}.
\begin{table}[!htb]
\begin{center}
\begin{tabular}{|c|c|c|c|c|c|c|c|c|}
\hline
$r_0, 1/yr$ & $q_0, 1/yr$ & $\sigma_0, USD/\sqrt{yr}$ & $r_k, 1/yr$ & $\sigma_k, 1/yr$ & $T, yr$   \\
\hline
0.02 & 0.03 & 2 $K$ & 0.01  & 5 & 1 \\
\hline
\end{tabular}
\caption{Parameters of the test.}
\label{tab1}
\end{center}
\end{table}
Let us recall, that here $\sigma(t)$ is the normal volatility. Therefore, we choose its typical value by multiplying the log-normal volatility by the strike level $K$. Also, for the values of parameters given in Tab.~\ref{tab1}, the last term practically vanishes, so in what follows we neglect it. Accordingly, from \eqref{indV}, \eqref{fDef} we find
\begin{align} \label{kT}
g(t) &= \exp \left[q_0 (t-T) + \frac{r_0}{r_k}\left( e^{- r_k t} - e^{- r_k T}\right) \right], \quad
f(x,t) = \frac{r_0}{r_k} \left(e^{- r_k T} - e^{- r_k t} \right).
\end{align}

\begin{figure}[!htb]
\begin{center}
\subfloat[]{\includegraphics[width=0.54\textwidth]{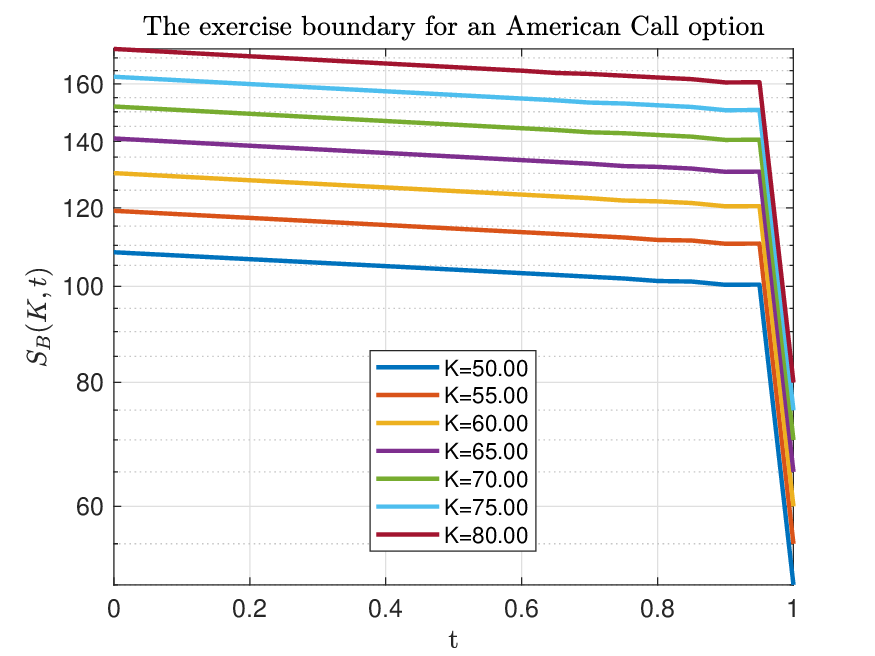}}
\hspace*{-0.3in}
\subfloat[]{\includegraphics[width=0.54\textwidth]{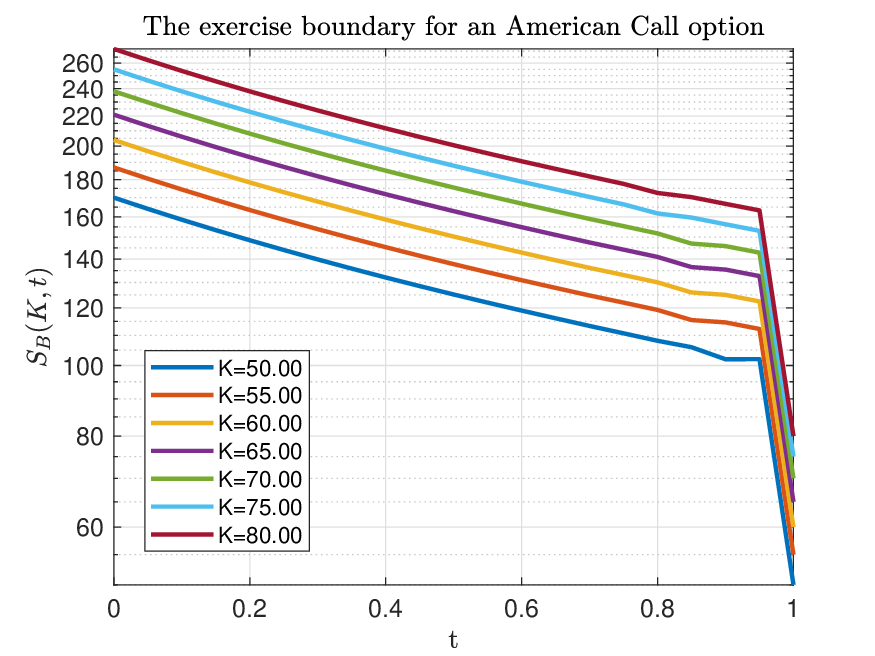}}
\end{center}
\caption{A semi-log plot of the exercise boundary $S_B(t)$ of the American Call option computed in the numerical example for various strikes $K \in [50:80]$; (a) parameters of the test are given in Table~\ref{tab1}; (b) parameters of the test are given in Fig.~\ref{EB}. }
\label{OUtest}
\end{figure}

 We run the test for a set of strikes $K \in [50, 55, 60, 65, 70, 75, 80]$. In Fig.~\ref{OUtest}a the exercise boundaries for the American Call option computed by solving \eqref{finVolterra2} using the trapezoid quadratures are presented as a function of the time $t$. For the initial guess the already computed value $y(t_{i-1})$ can be chosen, and the method typically converges within 3-4 iterations. The total elapsed time to compute $S_B(t)$ on a temporal grid: $t_i  = i \Delta t, \ i \in [0,M], \ \Delta t = T/M$ for all strikes with $M=20$ is 0.33 sec  in Matlab (fsolve) using two Intel Quad-Core i7-4790 CPUs, each 3.80 Ghz. Despite our Matlab code can be naturally vectorized, we didn't do it since in this example we used a simple method which can be approved in many different ways, i.e., see below in this paper. Also, it is worth mentioning that Matlab neither contain an embedded code for the Jacobi theta function (but it can be found in a third-party file exchange), nor for the derivatives of the theta function. Therefore, we refined translation of the original Pascal procedure made in the AMath library, see \citep{Elfun18, Fenton1984} and references therein, and equipped it with computation of the function derivatives. However, these functions are available in other languages, e.g., as a part of the python package mpmath,  \citep{mpmath} or in Wolfram Mathematica.

 Fig.~\ref{OUtest}b shows the results of the same test where parameters of the test are same as in Fig.~\ref{EB} (so, $r_K = \sigma_k = 0$. It can be seen that the exercise boundary is close to that presented in Fig.~\ref{EB}. The difference can be attributed to a difference in models (the OU vs the Black-Scholes model) because in this test the OU normal volatility $\sigma_{OU}$ is constant, and in the Black-Scholes model it is $\sigma_{BS} S$, i.e. a function of $S$, despite $\sigma_{OU} = \sigma_{BS} S$. Also, the elapsed time in the test at Fig.~\ref{EB} is 0.05 sec per one strike, i.e. this is 0.35 sec for 7 strikes as in the test in this Section. Thus, the elapsed time in both tests is very close, despite here we used our semi-analytical method, and for the test in Fig.~\ref{EB} - the trinomial tree. Note, that for the trinomial tree the number of discrete stock values was equal to the number of discrete time values which is 400, otherwise the accuracy of calculations is insufficient.

\section{One-factor models that can be reduced to the Bessel equation} \label{secBess}

In this section we consider two typical examples of one-factor time-dependent models such that the American option pricing can be reduced to solving the Bessel equation at the time-dependent spatial domain. Again, we chose pricing of the American Call option written on a stock which follows the time-dependent CEV model, and pricing of the American Put option written on a zero-coupon bond where the underlying interest rate follows the time-dependent CIR model.

The pricing problem for barrier options where the underlying follows the CIR or CEV model was considered in \citep{CarrItkinMuravey2020}. The authors describe two approaches to solve it: a) generalization of the method of heat potentials for the heat equation to the Bessel process, (so we call it the method of Bessel potentials), and b) the GIT method also extended to the Bessel process. In all cases, a semi-analytic solution was obtained meaning that first, one needs to solve a linear Volterra integral equation of the second kind, and then  the option price is represented as a one-dimensional integral. It was shown that computationally these methods are more efficient than both the backward and forward finite difference methods while providing better accuracy and stability. In more detail, see also \citep{ItkinLiptonMuraveyBook}.
Here we use same methodology but for pricing American rather than barrier options.

\subsection{The CEV model} \label{SecBO}

The time-dependent constant elasticity of variance (CEV) model is a one-dimensional diffusion process that solves a stochastic differential equation (SDE)
\begin{equation} \label{CEV}
d S_t = \mu(t) S_t dt + \sigma(t) S_t^{\beta+1} dW_t, \qquad S_{t=0} = S_0.
\end{equation}
Here $\mu(t)$ is the drift (under risk-neutral measure $\mu(t) = r(t) - q(t)$ where $r(t)$ is the deterministic short interest rate, and $q(t)$ is the continuous dividend), and $\beta$ is the elasticity parameter such that $\beta < 1, \ \beta \ne \{0, -1\}$\footnote{In case $\beta = 0$ this model is the Black-Scholes model, while for $\beta = -1$ this is the Bachelier, or time-dependent Ornstein-Uhlenbeck (OU) model.}. The model with constant coefficients was first introduced in \citep{c75} as an alternative to the geometric Brownian motion for modeling asset prices. We assume that all parameters of the model are known either as a continuous functions of time $t \in [0,\infty)$, or as a discrete set of $N$ values for some moments $t_i, \ i=1,\ldots,N$.

For the standard CEV process with constant parameters the change of variable  $z_t = 1/(\sigma|\beta|)S_t^{-\beta}$ transforms the SDE  \eqref{CEV} without  drift  ($\mu = 0$)  to  the  standard Bessel process of order $1/2 \beta$, \citep{RevuzYor1999, DavidovLinetsky2001}). As shown in \citep{CarrItkinMuravey2020}, a similar connection to the Bessel process can be established for the time-dependent version of the model in \eqref{CEV}. The authors consider an Up-and-Out barrier Call option written on the underlying process $S_t$. By the same standard argument, \citep{ContVolchkova2005, klebaner2005} its price $C(t,S)$ solves a parabolic PDE
\begin{equation} \label{PDEcev}
\fp{C}{t} + \dfrac{1}{2}\sigma^2(t) S^{2 \beta + 2}\sop{C}{S} +  [r(t) - q(t)] S \fp{C}{S} = r(t) C., \qquad
(t, S) \in [0, T] \times [0, S_B(t)].
\end{equation}

By analogy already established in Section~\ref{tdOU}, the price $C(t,S)$ of the American Call option written on the same underlying, in the continuation region solves the same PDE \eqref{PDEcev}. This equation should be solved subject to the terminal condition \eqref{tc0} and boundary conditions in \eqref{bc0}. Again, we first solve this pricing problem semi-analytically assuming that the exercise boundary $S_B(t)$ is known, and then derive a nonlinear integral Volterra equation for $S_B(t)$ that needs to be solved numerically.

As it was already emphasize, $S_B(t)$ in the pricing problem for the American Call option corresponds to the upper barrier $H(t)$ for the Up-and-Out option, while the boundary conditions at the moving boundary differ, namely: for the American Call option we have $C(t, S_B(t)) = S_B(t) - K$, while for the Up-and-Out barrier option this is $C(t, H(t)) = 0$ (if no rebate at hid is paid).

As shown in \citep{CarrItkinMuravey2020}, by a set of transformations the problem \eqref{PDEcev}, \eqref{tc0}, \eqref{bc0} can be transformed to the Bessel PDE
\begin{proposition}[Proposition 1.1 of \citep{CarrItkinMuravey2020}] \label{prop1}
The PDE in \eqref{PDEcev} can be transformed to
\begin{equation} \label{Bess}
\fp{u}{\tau} =\frac{1}{2} \sop{u}{z} + \frac{b}{z} \fp{u}{z},
\end{equation}
\noindent where $b$ is some constant, $u = u(\tau, z)$ is the new dependent variable, and $(\tau, z)$ are the new independent variables. The \eqref{Bess} is the PDE associated with the one-dimensional Bessel process, \citep{RevuzYor1999}
\begin{equation} \label{BesProc}
d X_t = d W_t  + \frac{b}{X_t} dt.
\end{equation}
\end{proposition}
Here the transformation is done in two steps. The first change of variables reads
\begin{equation} \label{tr1}
S = \left(-x \beta \right)^{-1/\beta}, \qquad C(t,S) \to u(t,x) e^{\int_T^t r(k) d k}, \qquad
\phi = \int_t^T \sigma^2(k) dk,
\end{equation}
\noindent and then the other change of variables is
\begin{align} \label{tr2}
z &= x F(\phi), \quad \tau  = \int_0^{\phi(t)} F^2(k) dk, \quad F(\phi) = e^{\int_0^\phi f(k) dk},
\quad f(t) = \beta \frac{r(t) - q(t)}{\sigma^2(t)}, \quad b = \frac{\beta + 1}{2\beta}.
\end{align}

The PDE in \eqref{Bess} should be solved in the domain $(\tau, z) \in [0, \tau(0)] \times [0, y(\tau)]$ subject to the terminal condition (compare with \eqref{tcFin})
\begin{equation} \label{tc0cev}
u(0, z) = 0,
\end{equation}
\noindent and the boundary conditions
\begin{equation} \label{bc0cev}
u(\tau,0) = 0, \qquad u(\tau,y(\tau)) = \left[ y(\tau)^{-1/\beta} \left( \frac{ \beta}{F(\phi(\tau))}\right)^{-1/\beta} - K \right] e^{-\int_0^{t(\tau)} r(k) d k}.
\end{equation}

Another important value is the option Delta at the moving  boundary $\Psi(\tau, z(\tau)) = \fp{u(\tau,z)}{z}|_{z = z(\tau)}$ which can be found explicitly by using Proposition~\ref{prop1}
\begin{equation} \label{grCEV}
\Psi(\tau, y(\tau))  =\frac{1}{\beta} \left(\frac{F(\phi(\tau))}{\beta}\right)^{1/\beta}  e^{-\int_0^{t(\tau)} r(k) d k}
[- y(\tau)]^{-1 -1/\beta}.
\end{equation}

As mentioned in \citep{CarrItkinMuravey2020}, the above formulae are valid at  $-1 < \beta < 0$. However, if $0 < \beta < 1$, the left boundary goes to $-\infty$. Therefore, in this case it is convenient to redefine $x \to \bar{x} = -x$. This also redefines $z \to \bar{z} = -z$. Then the domain of definition for $\bar{z}$ becomes $\bar{z} \in [\bar{y}(\tau), \infty),\ \bar{y}(\tau) = - y(\tau)$. One can also observe that by changing the sign of the strike $\bar{K} = -K$  we obtain a pricing problem for the American Put option.  The initial condition for this problem remains the same as in \eqref{tc0cev}, the boundary condition at the moving boundary - same as in \eqref{bc0cev}, but with the change $K \to \bar{K}, \ y(\tau) \to \bar{y}(\tau)$. And, as usual, a vanishing boundary condition for the Put option at $z \to \infty$ is natural to be set as well. Also, in case $0 < \beta < 1$ the PDE in \eqref{Bess} keeps the same form in the $\bar{z}$.

\subsection{The CIR model} \label{sCIR}

As applied to pricing barrier options the time-dependent Cox–Ingersoll–Ross (CIR) model was considered in \citep{CarrItkinMuravey2020}, so below we extract the main facts about this model from that paper. The original (time-independent) CIR model has been invented in \citep{cir:85} for modeling interest rates. In our time-dependent settings the CIR instantaneous interest rate $r_t$ follows the stochastic differential equation (SDE)
\begin{equation} \label{CIR}
d r_t = \kappa(t)[\theta(t) - r_t] dt + \sigma(t)\sqrt{r_t} dW_t, \qquad r_{t=0} = r.
\end{equation}
Here $\kappa(t) > 0$ is the speed of mean-reversion, $\theta(t)$ is the mean-reversion level. This model eliminates negative interest rates if the Feller condition $ 2 \kappa(t) \theta(t)/\sigma^2(t) > 1$ is satisfied, while still preserves tractability, see e.g., \citep{andersen2010interest} and references therein.

Since the CIR model belongs to the class of exponentially affine models, the price of the ZCB $F(r,t,Q)$ for this model is known in closed form. It is known, that $F(r,t,Q)$ under a risk-neutral measure solves a linear parabolic partial differential equation (PDE), \citep{privault2012elementary}
\begin{equation} \label{PDECIR}
\fp{F}{t} + \dfrac{1}{2}\sigma^2(t) r \sop{F}{r} + \kappa(t) [\theta(t) - r] \fp{F}{r} = r F,
\end{equation}
\noindent which should be solved subject to the terminal condition
\begin{equation} \label{termCIR}
 F(r,Q,Q)  = 1,
\end{equation}
\noindent and the boundary condition
\begin{equation} \label{bcCIR}
F(r,t,Q)\Big|_{r \to \infty} = 0.
\end{equation}
The second boundary condition is necessary in case the Feller condition is violated, so the interest rate $r_t$ can hit zero. Otherwise, the PDE in \eqref{PDECIR} itself at $r=0$ serves as the second boundary condition.

The ZCB price can be obtained from \eqref{PDECIR} assuming that the solution is of the form
\begin{equation} \label{affSol1}
F(r,t,Q) = A(t,Q) e^{ B(t,Q) r},
\end{equation}
\noindent where $A(t,Q), B(t,Q)$ solve the system
\begin{align} \label{equAff}
 \fp{B(t,Q)}{t} &= 1 + \kappa(t) B(t,Q) - \frac{1}{2} \sigma^2(t) B^2(t,Q), \quad B(Q,Q) = 0,\\
 \fp{A(t,Q)}{t} &= - A(t,Q) B(t,Q)  \theta (t) \kappa (t), \quad A(Q,Q) = 1. \nonumber
\end{align}
The first equation in \eqref{equAff} is the Riccati equation. It this general form it cannot be solved analytically for arbitrary functions $\kappa(t),  \sigma(t)$, but can be efficiently solved numerically. Also, in some cases it can be solved approximately (asymptotically), see e.g., an example in \citep{CarrItkin2020jd}.  Once the solution is obtained, the second equation in \eqref{equAff} can be solve analytically to yield
\begin{equation}
A(t,Q) = e^{- \int_Q^t B(m) \theta (m) \kappa (m) \, dm}.
\end{equation}
When coefficients $\kappa(t), \theta(t), \sigma(t)$ are constants, it is known that the solution $B(t,Q)$ can be obtained in closed form and reads, \citep{andersen2010interest}
\begin{equation}
B(t,Q) = - \frac{2[\exp((Q-t)h) - 1]}{2h + (\theta + h)[\exp((Q-t)h) - 1]}, \qquad h = \sqrt{\theta^2 + 2 \sigma^2}.
\end{equation}
Thus, $B(t,Q) < 0$ if $t < Q$. Therefore,  $F(r,t,Q) \to 0$ when $r \to \infty$. In other words, the solution in \eqref{affSol1} satisfies the boundary condition at $r \to \infty$. In case when all the parameters of the model are deterministic functions of time, and $B(t,Q)$ solves the first equation in \eqref{equAff}, this also remains to be true, \citep{CarrItkinMuravey2020}.

Our focus here is on pricing an American Put option written on a ZCB. Similar to Section~\ref{tdHW} and \citep{CarrItkinMuravey2020} we observe that under a risk-neutral measure the option price $C(t,r)$ in the continuation region solves the same PDE as in  \eqref{PDECIR}, \citep{andersen2010interest}.
\begin{equation} \label{PDEPCIR}
\fp{C}{t} + \dfrac{1}{2}\sigma^2(t) r\sop{C}{r} + \kappa(t) [\theta(t) - r] \fp{C}{r} = r C, \quad (t, r) \in [0,T] \times [r(t), \infty),
\end{equation}
\noindent where $r_B(t)$ is the exercise boundary. The terminal condition at the option maturity $T \le Q$ for this PDE reads
\begin{equation} \label{tc0CIR}
C(T,r) = \left(K - F(r,T,S)\right)^+ = 0,
\end{equation}
\noindent and the boundary conditions are same as in \eqref{bc1hw}, \eqref{bc2hw}.

\subsubsection{Reduction to the Bessel equation} \label{DOB}

The PDE in \eqref{PDEPCIR} can also be transformed to that for the Bessel process in \eqref{BesProc}.

\begin{proposition}[Proposition 2.1 of \citep{CarrItkinMuravey2020}]  \label{prop2}
The \eqref{PDEPCIR} can be transformed to
\begin{equation} \label{Bess1}
\fp{u}{\tau} = \frac{1}{2} \sop{u}{z} + \frac{b}{z} \fp{u}{z},
\end{equation}
\noindent where $b = m - 1/2$ is some constant, $u = u(\tau, z)$ is the new dependent variable, and $(\tau, z)$ are the new independent variables, if
\begin{equation} \label{cond1}
\frac{\kappa(t) \theta(t)}{\sigma^2(t)} = \frac{m}{2},
\end{equation}
\noindent where $m \in [0,\infty)$ is some constant. The \eqref{Bess1} is the PDE associated with the one-dimensional Bessel process in \eqref{BesProc}.
\end{proposition}
Here the following transformations were used
\begin{align} \label{trCIR1}
C(t,r) &= u(\tau,z) e^{a(t) r + \int_T^t a(s) \kappa(s) \theta(s) ds}, \qquad z = g(t) \sqrt{r}, \qquad t = t(\tau)\\
g(t) &=  \exp\left[ \frac{1}{2} \int_0^t \left( \kappa(s) - a(s) \sigma^2(s) \right) \, ds \right], \qquad
\tau(t) = \frac{1}{4} \int_t^T  g^2(s) \sigma^2(s) \, ds, \nonumber
\end{align}
\noindent where $a(t)$ solves the Riccati equation
\begin{equation} \label{ric2}
\frac{da(t)}{dt} = -\frac{\sigma^2(t) a^2(t)}{2} + \kappa(t) a(t)  + 1.
\end{equation}
The function $t(\tau)$ is the inverse map defined by the last equality in \eqref{trCIR1}. It can be computed for any $t \in [0,T]$ by substituting it into the definition of $\tau$, then finding the corresponding value of $\tau(t)$, and inverting.

As follows from Proposition~\ref{prop2}, for the time-dependent CIR model the transformation from \eqref{PDEPCIR} ro \eqref{Bess1} cannot be done unconditionally. However, from practitioners' points of view the condition \eqref{cond1} seems not to be too restrictive, Indeed, the model parameters already contain the independent mean-reversion rate $\kappa(t)$ and volatility $\sigma(t)$.  Since $m$ is an arbitrary constant, it could be calibrated to the market data together with $\kappa(t)$ and $\sigma(t)$. Therefore, in this form the model should be capable for calibration to the term-structure of interest rates, again in more detail see \citep{CarrItkinMuravey2020}.

In new variables the terminal and boundary conditions for \eqref{Bess1} read
\begin{align} \label{tc0CIRu}
u(0,z) &= 0, \qquad  u(\tau,z)\Big|_{z \to \infty} =0,  \qquad
u(\tau, y(\tau))  = K - A(t,Q) e^{B(t,Q) g^{-2}(t) y^2(\tau)},  \qquad t = t(\tau).
\end{align}

\subsection{Nonlinear integral Volterra equation for $y(\tau)$} \label{nonLinVoltCEV}

Let us, for example,  consider the CEV model with $-1 < \beta < 0$, so by the Proposition~\ref{prop1} the pricing PDE in the continuation region can be transformed to the Bessel PDE \eqref{Bess} with  a homogeneous  terminal condition \eqref{tc0cev} and the boundary conditions \eqref{bc0cev}.  Let us compare this problem with a similar one for the time-dependent OU model which was solved in Section~\ref{voltOU}. It can be observed that both problems look similar, namely: they both have a homogeneous terminal condition and the boundary condition at $x=0$, and the other boundary condition at the moving boundary $y(\tau): u(\tau, y(\tau)) = f^+(\tau)$, where $f^+(\tau)$ is some continuous function  of $\tau$. Thus, those two problems differ only by the PDE itself: in for the OU model this is a heat equation \eqref{Heat} while for the CEV model this is a Bessel equation \eqref{Bess}.

Based on our result obtained in \citep{CarrItkin2020jd} and replicated in \eqref{uFourier}, the solution of \eqref{Heat} with the corresponding terminal and boundary conditions  can be represented via the following ansatz
\begin{align} \label{genSol}
u(\tau, x) &= \int_0^\tau \Psi(s, y(s)) B_0(x,\tau \,|\, y(s), s) + \int_0^\tau \int_{0}^{y(s)} \lambda(s, \xi) B_0(x,\tau \,|\, \xi, s)  d\xi ds + \frac{x}{y(\tau)} f^+(\tau), \\
B_0(x,\tau \,|\, \xi, s) &= \frac{1}{2 y(\tau)}\left[ \theta_3(\phi_-(x,\xi,\tau), \omega_2(s, \tau)) - \theta_3(\phi_+(x,\xi,\tau),\omega_2(s, \tau)) \right]. \nonumber
\end{align}
Here $\theta_3(\phi_-(x,\xi,\tau),\omega_2(s, \tau))$ and $\theta_3(\phi_+(x,\xi,\tau),\omega_2(s, \tau))$ are two {\it periodic} solutions of the heat equation in "conjugate" variables \footnote{They are Green's functions of this problem at the finite interval $x \in [0, y(\tau)]$ with the initial condition at $x=\xi$ for $\theta_3(\phi_-(x,\xi,\tau),\omega_2(s, \tau))$ and $x = - \xi$ for $\theta_3(\phi_+(x,\xi,\tau),\omega_2(s, \tau))$.}, \citep{mumford1983tata}
\begin{equation} \label{HeatTheta}
\fp{}{s} \theta_3(\phi_\pm(x,\xi,\tau),\omega_2(s, \tau)) = -\frac{1}{4} \sop{}{\xi} \theta_3(\phi_\pm(x,\xi,\tau),\omega_2(s, \tau)).
\end{equation}
The difference of two Green's functions in the definition of $B_0(x,\tau \,|\, y(s), s)$ is necessary to obey the boundary conditions (like in the method of images, \citep{TS1963}.

Constructing the solution in such a way and explaining this construction in such a form is useful because it allows a natural generalization of this approach. In particular, as applied to the problem in \eqref{PDEcev} with the terminal and boundary conditions in \eqref{tc0cev}, \eqref{bc0cev}, by analogy one can claim that the solution of this problem is also given by \eqref{genSol}, where now a {\it periodic} solution of the heat equation has to be replaced with a {\it periodic} solution of the Bessel equation.

Fortunately,  such a periodic solution has been already constructed in \citep{CarrItkinMuravey2020, ItkinMuraveySabrJD}. In that paper we introduced a new function which we call {\it the Bessel Theta function}
\begin{equation} \label{thetaBess}
\Theta_{\nu}(\theta, x_1, x_2)   = \frac{2}{y(\tau)} (x_1 x_2)^{|\nu|} \sum_{n = 1}^\infty e^{- \mu_n^2 \theta^2}  \frac{\Jnum( \mu_n x_1) \Jnum(\mu_n x_2)}{J_{|\nu| + 1}^2 (\mu_n)}.
\end{equation}
Here $J_\nu(x)$ is the Bessel function of the first kind, $\nu = 1/(2\beta) < 0$, since $\beta < 0$, $\mu_n$ is an ordered sequence of positive zeros of the Bessel function $\JnuM(\mu)$:
\[
J_{|\nu|}(\mu_n) = J_{|\nu|}(\mu_m) = 0, \quad \mu_n > \mu_m > 0, \quad n >m.
\]
The function $\Theta_{\nu}(\theta, x_1, x_2) $ is an analog of the Jacobi theta function (which is a periodic spatial solution of the heat equation) in case of the Bessel equation.

It can be seen that at $\beta = -1,  \nu = -1/2$ and $|\nu| = 1/2$, our CEV problem becomes a time-dependent OU problem considered in Section~\ref{voltOU}. Therefore, taking into account that at $\nu = -1/2$, we have, \citep{as64}
\begin{align*}
J_{-\nu}(x) &= \sqrt{\frac{2}{\pi x }} \sin (x), \qquad \mu_n = n \pi, \ n=0,1,... \\
\Theta_{\nu} &\left(\frac{\sqrt{\tau-s}}{y(\tau)}, \frac{y(s)}{y(\tau)}, \frac{x}{y(\tau)} \right) =
\frac{2}{y(\tau)}  \sqrt{\frac{y(s)}{y(\tau )} \frac{x}{y(\tau )}}  \sum_{n = 1}^\infty e^{-\frac{\mu_n^2 (\tau-s)}{y^2(\tau)}}  \frac{\Jnum(\mu_n y(s)/y(\tau)) \Jnum(\mu_n x/y(\tau))}{J_{|\nu| + 1}^2 (\mu_n)} \\
&= \frac{2}{y(\tau)} \sum_{n = 1}^\infty e^{-\frac{\mu_n^2 (\tau-s)}{y^2(\tau)}} \frac{\pi ^2 n \sqrt{\frac{n y(s)}{y(\tau )}} \sqrt{\frac{n x}{y(\tau )}} \sin \left(\frac{\pi  n y(s)}{y(\tau )}\right) \sin \left(\frac{\pi  n x}{y(\tau )}\right)}{[\sin (\pi  n)-\pi  n \cos (\pi  n)]^2 \sqrt{\frac{x y(s)}{y(\tau )^2}}} \\
&= \frac{1}{y(\tau)} \left[ \sum _{n=1}^{\infty } \omega_2(s, \tau)^{n^2} \cos \left(\frac{\pi  n (x-y(s))}{y(\tau )}\right) - \sum _{n=1}^{\infty } \omega_2(s, \tau)^{n^2} \cos \left(\frac{\pi  n (y(s)+x)}{y(\tau )}\right) \right] \\
&=  \frac{1}{2 y(\tau)} \left[ \theta_3(\phi_-(x,\xi,\tau),\omega_2(s, \tau)) - \theta_3(\phi_+(x,\xi,\tau),\omega_2(s, \tau)) \right] = B_0(x,\tau \,|\, \xi, s).
\end{align*}
Therefore, in this limit we restore the correct solution of \eqref{Heat} given in \eqref{uFourier}.

\begin{figure}
\centering
\includegraphics[totalheight=2.7in]{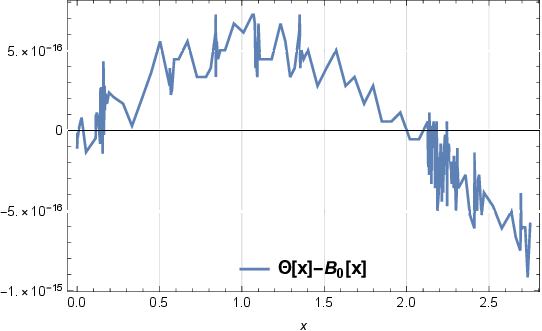}
\cprotect\caption{Convergence of $Theta_\nu(x)$ with $\nu = -0.5, \ s = 0.5, \ \tau = 1$, the sum is over $n \in [-M, M], \  M=10$, and $B_0(x,\tau \,|\, y[s], s)$ is computed in Wolfram Mathematica by using function \verb|EllipticTheta(3,z,q)|.}
\label{FigConv}
\end{figure}

The sum in the definition number of $\Theta_\nu(...)$ usually quickly converges due to the exponential term. For instance, in Fig.~\ref{FigConv} a difference between $\Theta_{\nu} \left(\frac{\sqrt{\tau-s}}{y(\tau)}, \frac{y(s)}{y(\tau)}, \frac{x}{y(\tau)} \right)$ and $B_0(x,\tau \,|\, \xi, s)$ is presented by using a test function $y(s) = 5 - \left(\frac{1}{2} + s\right)^2$, and the values $s = 0.5, \tau = 1$, and the sum is over $n \in [-M, M], \, M=10$.

\begin{figure}
\centering
\fbox{\includegraphics[width=\textwidth]{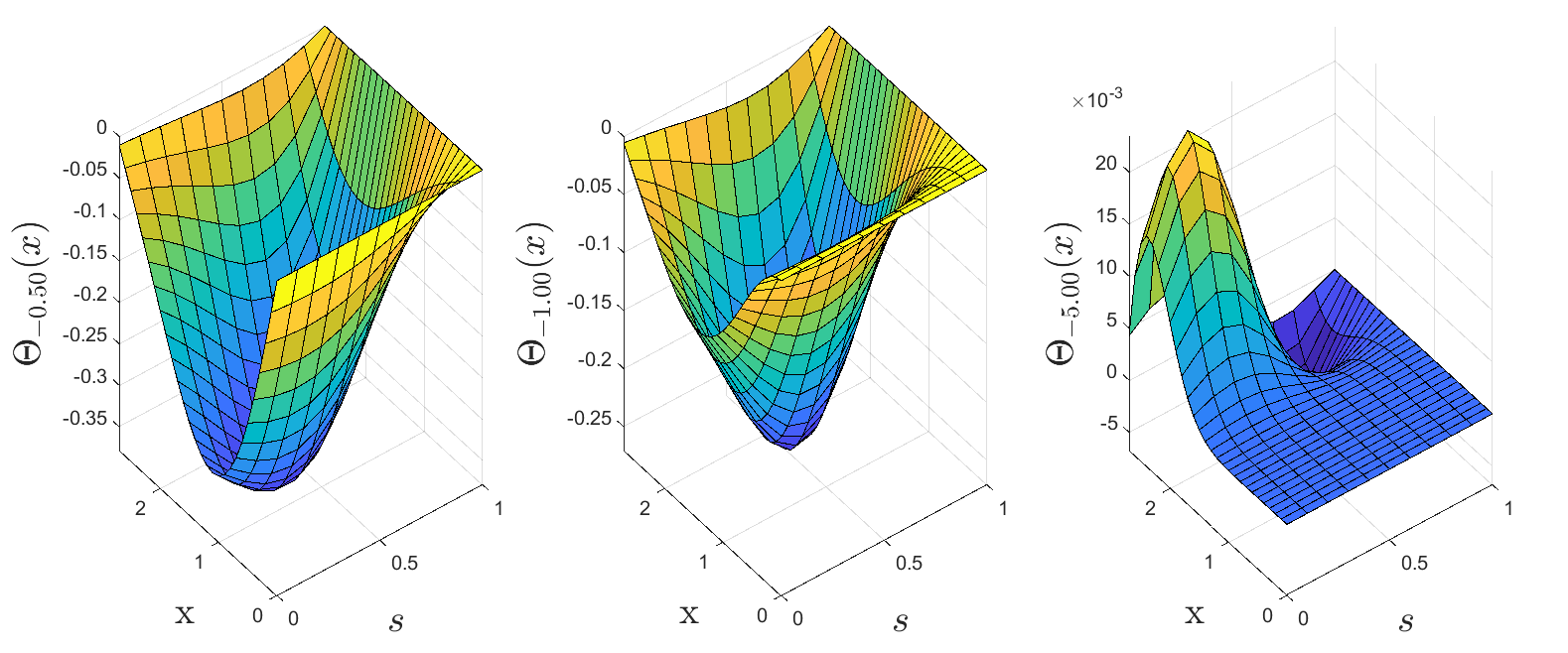}}
\caption{3D plot of $\Theta_{\nu} \left(\frac{\sqrt{\tau-s}}{y(\tau)}, \frac{y(s)}{y(\tau)}, \frac{x}{y(\tau)} \right)$ as a function of $(x, s)$ obtained in the same experiment for three values of $\nu = 1/(2 \beta)$: $\beta = -1, -1/2, -0.1$.}
\label{FigTheta}
\end{figure}

In Fig.~\ref{FigTheta} we also present a 3D plot of $\Theta_{\nu} \left(\frac{\sqrt{\tau-s}}{y(\tau)}, \frac{y(s)}{y(\tau)}, \frac{x}{y(\tau)} \right)$ as a function of $(x, s)$ obtained in the same experiment for 3 values of $\nu = 1/(2 \beta)$: $\beta = -1, -1/2, -0.1$.

Thus, using this analogy, we can represent the solution of \eqref{Bess} as
\begin{align}  \label{uBessFourier}
u(\tau, x) &=  \frac{x}{y(\tau)} f^+(\tau) + \int_0^\tau \Psi(s, y(s)) \Theta_{\nu} \left(\frac{\sqrt{\tau-s}}{y(\tau)}, \frac{y(s)}{y(\tau)}, \frac{x}{y(\tau)} \right) ds \\
&+ \int_0^\tau \int_{0}^{y(s)} \lambda(s, \xi) \Theta_{\nu} \left(\frac{\sqrt{\tau-s}}{y(\tau)}, \frac{\xi}{y(\tau)}, \frac{x}{y(\tau)} \right)  d\xi ds.  \nonumber
\end{align}
Here the function $f^+((s)$ is given by \eqref{bc0cev} and reads
\begin{equation}
f^+((s) = \left[ y(s)^{-1/\beta} \left( \frac{ \beta}{F(\phi(s))}\right)^{-1/\beta} - K \right] e^{-\int_0^{t(s)} r(k) d k}.
\end{equation}
The gradient $\Psi(s, y(s))$ is also known explicitly and is given in \eqref{grCEV}.

Accordingly, the nonlinear integral Volterra equation for the exercise boundary $y(\tau)$ has the same form as \eqref{finVolterra1}, where now we need to replace the theta function $\theta_3\left( \phi_-(y(\tau), y(s), \tau), \omega_2(s, \tau) \right)$ and its derivatives with
\begin{align*}
\frac{1}{y(\tau)} \theta_3\left( \phi_-(y(\tau), y(s), \tau), \omega_2(s, \tau) \right) &\mapsto
 \Theta_{-\nu} \left(\frac{\sqrt{\tau-s}}{y(\tau)}, \frac{y(s)}{y(\tau)}, 1 \right) +
\Theta_{\nu} \left(\frac{\sqrt{\tau-s}}{y(\tau)}, \frac{y(s)}{y(\tau)}, 1 \right)  + 1, \nonumber \\
\frac{\pi}{2 y(\tau)^2} \theta'_3\left( \phi_-(y(\tau), y(s), \tau), \omega_2(s, \tau) \right) &\mapsto \frac{\partial}{\partial z} \Bigg\{ \Theta_{-\nu} \left(\frac{\sqrt{\tau-s}}{y(\tau)}, \frac{y(s)}{y(\tau)}, z \right) +
\Theta_{\nu} \left(\frac{\sqrt{\tau-s}}{y(\tau)}, \frac{y(s)}{y(\tau)}, z \right) \Bigg\}\Bigg|_{z = 1}.
\end{align*}
\vspace{-\baselineskip}
\begin{align}
B_1(y(\tau),\tau \,|\, \xi, s)  &\mapsto \frac{1}{y(\tau)}\frac{\partial}{\partial z} \Theta_{\nu} \left(\frac{\sqrt{\tau-s}}{y(\tau)}, \frac{\xi}{y(\tau)}, z \right)\Bigg|_{z = 1} \\
&= 2 \xi^{|\nu|} y^{|\nu|-2}(\tau) \sum_{n = 1}^\infty e^{- \frac{\mu^2_n (\tau-s)}{y(\tau )^2}}
\frac{J_{-\nu}\left(\frac{\mu_n \xi}{y(\tau)} \right)} {J_{|\nu|+1}(\mu_n)^2}
\Big[ (|\nu|+\nu) J_{-\nu}(\mu_n) + \mu_n  J_{-\nu-1}(\mu_n) \Big], \nonumber \\
\end{align}
This yields
\begin{align} \label{finVolterraCEV}
\Psi(\tau, y(\tau)) &=  \frac{1}{y(\tau)} f^+(\tau) + \int_0^\tau \Psi(s, y(s)) B_1(y(\tau),\tau \,|\, y(s), s) ds
+ \int_0^\tau \int_{0}^{y(s)} \lambda(s, \xi) B_1(y(\tau),\tau \,|\, \xi, s) d\xi ds.
\end{align}
Since by \eqref{wEq}
\begin{equation*}
\lambda(\tau,x) = -x \frac{\partial}{\partial \tau} \left( \frac{f^+(\tau)}{y(\tau)}\right),
\end{equation*}
\noindent and from \eqref{bc0cev}
\begin{equation*}
f^+(\tau) = \left[ y(\tau)^{-1/\beta} \left( \frac{ \beta}{F(\phi(\tau))}\right)^{-1/\beta} - K \right] e^{-\int_0^{t(\tau)} r(k) d k}
\end{equation*}
\noindent the inner integral in the last line of \eqref{finVolterraCEV} can be further simplified, since
\begin{align*}
\int_0^{y(s)} & \xi^{|\nu|+1} J_{-\nu} \left(\frac{\mu_n \xi }{y(\tau )}\right) \, d\xi = \\
&- \frac{2^{\nu } y^{| \nu | +2}(s) }{[-|\nu| + \nu -2] \Gamma (1-\nu )}
\left(\frac{\mu_n^2 y(s)^2}{y(\tau )^2}\right)^{-\nu/2}\,  _1F_2\left(-\frac{\nu }{2}+\frac{| \nu | }{2}+1;1-\nu ,-\frac{\nu }{2}+\frac{| \nu | }{2}+2;-\frac{\mu _n^2 y(s)^2}{4 y(\tau )^2}\right),
\end{align*}
\noindent where $_1F_2\left(a_1; b_1, b_2;, z \right)$ is the generalized hypergeometric function, \citep{as64}.

This fully solves the problem under consideration. For the other interval $x \in [y(\tau), \infty)$ the solution can be constructed in the same way, and we leave it for the reader. Also, a method of solving a linear Volterra equation for the CEV and CIR models as well as computation of $\Theta_{\pm \nu}$ function are discussed in detail in \citep{CarrItkinMuravey2020}. Same approach can be used here for solving the corresponding nonlinear Volterra equation, also see the advanced discussion in the next Section.

\section{Solving nonlinear Volterra equations by various methods including ML} \label{ML}

In this Section we discuss various methods to solve non-linear non-standard integral Volterra equations of the type \eqref{VoltDef}. Traditional method for solving non-linear equations of this type could be conventionally split into three groups, see e.g., a recent paper \citep{Nedaiasl2019} and references therein.

If the expression under the integral (the non-linear integral kernel) doesn't have singularities, then {\it direct quadratures} can be used to approximate the integral term. The term $y'(s)$ under the integral can be approximated by using finite differences of the necessary order of approximation. Thus obtained non-linear algebraic equation for $y(s)$ can be solved numerically using any standard method (e.g., the Newton-Raphson method) sequentially starting from $s=0$ and the $s = 1,\ldots, M$. For doing so, one can build a discrete grid in time $\Omega: s \in [0, \tau]$ with $M+1$ nodes equally distributed at $[0, \tau]$ with the step $\delta = \tau/M)$. Then, instead of a continuous variable $y(s)$  we obtain a discrete vector $y_i, \ i=0,\ldots,M$, such that $y_0 = y(0) = K, \ y_{M} = y(\tau)$. Alternatively, a system of non-linear algebraic equations for the entire vector $y(0),\ldots,y(M)$ can be solved given some good initial guess.

Another possible form of this method can be constructed based on the observation that all the Volterra equations we need to solve, e.g., \eqref{finVolterra2}, are linear in $\Psi(\tau, y(\tau)$. Since the connection between $\Psi(\tau, y(\tau)$ and $y(\tau)$ is analytic (see, e.g, \eqref{uEB} or \eqref{PsiHW}), the algorithm, e.g., for \eqref{finVolterra2} could be as follows. Given the initial guess for $y(s), \ s=0,\ldots,M$, one can compute the kernels $\theta'_3\left( \phi_-(y(\tau),y(s), \tau), \omega_2(s, \tau) \right)$ and the second integral in the RHS. of \eqref{finVolterra2}. This gives rise to a {\it linear} integral Volterra equation of the second kind for $\Psi(\tau, y(\tau)$. Once the solution is found on the grid (by using any standard method), new values of $y_i, \ i=0,\ldots,M$ can be found by solving \eqref{uEB} w.r.t. to $y_i$. Then the iterations can be continued until necessary convergence is achieved.

In case the integral kernel contains some week (integrated) singularities, like in \eqref{finVoltalt} which contains week singularities $(\tau-s)^{-1/2}$, a barycentric interpolation method can be used to resolve this, so the integral can we well approximated on the grid with no singularities in the final solution, in more detail see \citep{Nedaiasl2019}.

Another class of methods that can be used for solving non-linear integral equation of the type \eqref{VoltDef} are {\it interpolation or collocation-based} methods. The idea of these methods is to approximate the continuous solution onto a finite-dimensional subspace. This can be done by using various polynomial or polynomial-spectral collocation procedures, e.g. see \citep{Andersen2016} among others. Note that the approach of \citep{Trefethen} implemented in the \verb|chebfun| package in Matlab, where piecewise polynomial interpolants and Chebyshev polynomials are used for functional representation of objects, also belongs to this class. The package aims to combine the feel of symbolic computing systems like Maple and Mathematica with the speed of floating-point numerics, and also includes an API for solving nonlinear Volterra equations with high precision.

In \citep{CarrItkinMuraveyHeston} a three-dimensional integral Volterra (linear) equation was solved by approximating the kernel with the Radial Basis Functions (RBF) as this was proposed in \citep{Assari2019,Zhang2014, ItkinMuraveySabrJD}  (see also references therein). In doing so, we proposed a new set of RBFs which allow computation of one integral (out of three) in closed form. It is proved that these new RBF can be used for interpolation and the corresponding analysis is provided in the paper. We then compare performance of our method with that of a modern FD approach and find that the former outperforms the later.

 A recent paper \citep{Lu2023} exploits the same idea but wrapped out into a ML framework. In more detail, the authors construct a new neural network (NN) method to solve linear Volterra and Fredholm integral equations based on the sine-cosine basis functions and extreme learning machine (ELM) algorithm. The NN consists of an input layer, a hidden layer, and an output layer, in which the hidden layer is eliminated by utilizing the sine-cosine basis functions. Furthermore, the problem of ﬁnding network parameters is converted into solving a set of linear equations.

 Looking at this method, it is clear that the choice of the basis is not limited to just sine and cosine functions. The best choice of the basis should be that one which allows computation of the integral in the Volterra equation in closed form. For linear equations and simple kernels this can  be done relatively easy. However, for nonlinear equations this could be either impossible at all (hence, instead the integral has to be computed numerically). or can be extremely bulky. Again, out trick of switching from non-linear equation for $y(\tau)$ to a linear one for $\Psi(\tau, y(\tau)$ given the current values of $y_i$ can help here, so this method can now be undoubtedly used, but it still requires iterations to converge. Once the expansion of $\Psi(\tau, y(\tau)$ into series of basis functions is found (all coefficients are computed), the value of $y(\tau)$ can also be easily computed by inverting (either analytically or numerically) the map $\Psi(\tau, y(\tau) \mapsto y(\tau)$.

The main advantages of interpolation-based methods lies in the fact that once interpolation coefficients are found, the approximate solution $y(s)$ can be obtained at any point $s \in [0, \tau(0)]$, in contrast to the direct quadratures method where the solution is known only in the grid points. A possible cost one has to pay for this is the speed of the method. But as usual for the ANN methods, the problem can be trained offline, so using the already trained ANN the solution can be obtained very fast.

For time-dependent models calibration of the model (training the ANN) could be a complex problem. Indeed, following \citep{ItkinDL2020} consider a financial model $\mathcal{M}$ with parameters of the model $p_1,\ldots,p_n, \ p_i \in \mathbb{R}, n \in \mathbb{Z}, n \ge 1$ which provides prices of some financial instruments given the input data $\theta_1,\ldots,\theta_l, \ l \in \mathbb{Z}, l \ge 1$. For example, the OU model described in Section~\ref{SecHeat} has three time-dependent parameters $r(t), q(t), \sigma(t)$, and three input parameters $S, K, T$. The ANN approach for option pricing basically assumes that the ANN can be used as a universal approximator $\mathbb{R}^{n+l}  \to \mathbb{R}$, i.e. given a vector of the input data $\bm{\theta}$ and a vector of the values of the model  parameters $\bm{p}$ it provides a unique option price, e.g., the Call option price $C(\bm{\theta},\bm{p})$. Since parameters of the model are time-dependent, these time dependencies have to be described either analytically (hence, then constant parameters of this dependencies become new (additional) parameters of the entire model), or by another ANN trained accordingly to the market data. Once this is done, the whole model can be trained either to the available market data (thus, the model calibration is embedded into this step), or to the "reference" solutions of the corresponding integral Volterra equation. The "reference" solution, which is an output of the training samples, can then be obtained by using the other (e.g., direct) methods. The trained ANN then can be used for pricing American options for both in-sample and out-of-sample data.

Perhaps, this approach is too expensive to apply it to simple one-dimensional problems when the number of Volterra equations to be solved is small. However, with the increase of dimensionality, e.g. for stochastic volatility models or for basket options, it could be optimal. However, then speed-wise it should be compared with a similar ML approach applied directly to solving a corresponding PDE in the continuation region.

\section{Conclusions}

In this paper we demonstrate how the GIT technique could be extended from semi-analytic pricing of barrier options, \citep{ItkinLiptonMuraveyBook} to that for American options. Using as example some one-factor time-dependent models for the underlying, we describe in detail how the corresponding non-linear integral Volterra equation for the exercise boundary can be derived for both Call and Put options. To the best of our knowledge, before this result was known only for the Black-Scholes model with constant coefficients. We also provide some examples how thus obtained non-linear Volterra equations can be solved numerically.

To underline, in the paper we obtain semi-analytic representations of American options prices for the time-dependent models where the pricing PDE can be reduced either to the Heat or Bessel equation with a general source term and moving boundaries. The corresponding solutions in the continuation region are obtained analytically by using the GIT technique, however, the final result can be also treated as  an application of the Duhamel's principle to the corresponding problem. We believe this is also an interesting contribution, since we didn't find in the literature any reference to using the Duhamel's formula for problems with moving boundaries.

Moreover, we provide a roadmap how other similar PDEs can be solved in the same manner. The idea, e.g., for the domain $x \in [0,y(\tau)]$, is that when pricing American options the terminal and boundary conditions are the same for any model and read $u(0,x) = u(\tau, 0) = 0, \ u(\tau, y(\tau)) = f^+(\tau)$, but the PDE themselves could differ. However, once a {\it periodic} solution of this PDE with fully homogeneous conditions (i.e., with $u(\tau, y(\tau)) = 0$ as well) is known, it can be plugged into the general ansatz of the solution given in \eqref{genSol} to provide the solution of the original problem. Same is true for the nonlinear integral Volterra equation solved by $y(\tau)$, which can be obtained in the same way, for instance,  from \eqref{finVolterra2}.

Since the GIT approach can also be applied to pricing options under various stochastic volatility models, \citep{ItkinMuraveySabrJD, CarrItkinMuraveyHeston}, by analogy it is expected that the same method can be used for American options as well. Currently, this is our work in progress.

\section*{Acknowledgments}

Andrey Itkin is very obliged to Peter Carr for numerous discussions on American options pricing in the past. We also thank Alex Lipton for discussions on a heat potential method

\section*{Disclaimer}

This paper represents the opinions of the authors. It is not meant to represent the position or opinions of the ADIA and NYU or their Members, nor the official position of any staff members. Any errors are our own.

%%%%%%%%%%%%%%%%%%%%%%%%%%%%%%%%%%%%%%%%%%%%%%%%%%%%%%%%%%%%
\vspace{0.4in}
%\printbibliography[title={References}]

%============== arxiv ============
%\bibliographystyle{apalike}
%\bibliography{tdOU}
%  Also commented out lines 212-290 in mydef2col.sty

\vspace{0.4in}
\appendixpage
\appendix
\numberwithin{equation}{section}
\setcounter{equation}{0}

\section{Duhamel's principle for the domain $x \in [0, y(\tau)]$} \label{app1}

Here we want to solve \eqref{wEq} and show, that the Duhamel's principle can be applied to problems with moving boundaries. For doing so, we use a general idea of the GIT method: first, construct a certain integral transform and solve thus obtained an odinary differential equation (ODE) for the image $\bar{u}$, and second, construct an inverse transform that finally solves the problem.

Let us consider a pair  of integral transforms $\bar u _\pm $
\begin{equation}  \label{GIT-strip}
	\bar u _\pm  (\tau, \omega)= \int_{0}^{y(\tau)} e^{\pm \iu \omega x } U(\tau, x) dx
\end{equation}
\noindent and apply them to \eqref{wEq}. We have
\begin{align*}
	\int^{y(\tau)}_{0} e^{\pm \iu \omega x } \sop{U(\tau, x)}{x} dx &=  \fp{U(\tau, x)}{x} \Big |_{x = y(\tau)} e^{\pm \iu \omega y(\tau) }-\fp{U(\tau, x)}{x} \Big |_{x = 0}
	\mp \iu \omega  	\int^{y(\tau)}_{0} e^{\pm \iu \omega x } \fp{V(\tau, x)}{x} dx \\
	&= \Psi(\tau, y(\tau)) e^{\pm \iu \omega y(\tau)} -\Phi(\tau, y(\tau))- \omega^2 \bar v _{\pm}, \\
	\int_{0}^{y(\tau)} e^{\pm \iu \omega x } \fp{U(\tau, x)}{\tau} dx &= \frac{d}{d \tau} \left[ 	\int_{0}^{y(\tau)} e^{\pm \iu \omega x } U(\tau, x) dx\right] - e^{\pm \iu \omega x} U(\tau, y(\tau)) y'(\tau)
	= \frac{d \bar u _\pm}{d \tau}.
\end{align*}

Here the functions $\Psi(\tau, y(\tau))$ and $\Phi(\tau, y(\tau))$ are  gradients of the solution at the  boundaries $x = y(\tau)$ and $x = 0$, respectively.
\begin{equation}
	\Psi(\tau, y(\tau)) =  \fp{U(\tau, x)}{x} \Big |_{x = y(\tau)}, \quad 	\Phi(\tau, y(\tau)) =  \fp{U(\tau, x)}{x} \Big |_{x =0}.
\end{equation}
Collecting all the terms together, we obtain the ODE for $\bar u _\pm $
\begin{align} \label{uODE}
	\frac{d \bar u_\pm (\tau, \omega)}{ d\tau}  +\omega^2 \bar u _\pm &= \Psi(\tau, y(\tau)) e^{\pm \iu \omega y(\tau)} - \Phi(\tau, y(\tau))  + \int^{y(\tau)}_{0} e^{\pm \iu \omega x } \lambda(\tau, x)dx,   \\
	\bar u(0, \omega) &= 0. \nonumber
\end{align}
Solving \eqref{uODE} by using a standard technique yields the explicit formula for $\bar u_\pm (\tau, \omega)$
\begin{align} \label{barv_explicit}
	\bar u _\pm(\tau, \omega) = \int_0^\tau e^{-\omega^2 (\tau -s)}
	\left[
	\Psi(s, y(s)) e^{\pm \iu \omega y(s) } - \Phi(s, y(s))+
	\int^{y(s)}_{0}  e^{\pm \iu \omega x }  \lambda(s, x)dx
	\right] ds. \nonumber
\end{align}

To the best of our knowledge, each transform $\bar u _+$ and $\bar u _-$ is not explicitly invertible, however their linear combination is. Indeed,
\begin{align*}
	\bar u(\tau ,\omega) = \frac{\bar u_+  - \bar u_- }{2 \iu} = \frac{1}{\iu} \int^{y(\tau)} _{0}  U(\tau, x )
	\sinh \left( \iu \omega x \right) dx
	= \int^{y(\tau)}_{0} U(\tau, x) \sin \left( \omega x \right) dx,
\end{align*}
\noindent which is a sine transform with a well-known inversion formula. On the other hand, the function $\bar u (\tau, \omega)$ can be represented as follows
\begin{align*}
	\bar u (\tau, \omega) &=   \int_0^\tau e^{-\omega^2 (\tau -s)}
	\left[
	\Psi(s, y(s)) \sin \left( \omega y(s)  \right)  +  \int_{y(s)}^{+\infty}  \lambda(s, x) \sin \left( \omega x \right)  dx
	\right] ds.
\end{align*}

Using the inversion formula for the sine transform
\begin{equation} \label{fourier_series_inversion}
	U(\tau, x) = \frac {2}{y(\tau)}  \sum_{n = 1}^{+\infty} \bar u(\tau, \pi n / y(\tau)) \sin \left( \frac{\pi n x}{ y(\tau)}\right) d\omega .
\end{equation}
\noindent we finally obtain
\begin{align} \label{UFinal}
	U(\tau, x) = \frac {2}{y(\tau)}  \sum_{n = 1}^{\infty}  \sin \left(\frac{\pi n x}{ y(\tau)}\right) \int_0^\tau e^{-\omega^2 (\tau -s)}
	\left[
	\Psi(s, y(s)) \sin \left( \frac{\pi n y(s)}{ y(\tau)}  \right)  +  \int_{0}^{y(s)}  \lambda(s, \xi) \sin \left( \frac{\pi n \xi}{ y(\tau)} \right)  d\xi
	\right] ds .
\end{align}

\section{Regularization of \eqref{U_final} } \label{app2}

A straightforward differentiation of \eqref{U_final} on $x$ yields
\begin{align*}
	\fp{U(\tau, x)}{x}  \Big|_{x = y(\tau) + 0}  &= \frac{1}{2\sqrt{\pi \tau}} \int_{y(0)}^{\infty} f(\xi)
	\left[
	-\frac{y(\tau) - \xi}{2\tau} e^{-\frac{(y(\tau) - \xi)^2}{4\tau}} + \frac{\xi - y(\tau)}{2\tau} e^{-\frac{(y(\tau) - \xi)^2}{4\tau}}
	\right] \\
	&+\int_{0}^\tau \frac{\Psi(s,y(s)) - y'(s) g(s)}{2\sqrt{\pi (\tau - s)^3}} (y(\tau) - y(s)) e^{-\frac{(y(\tau) - y(s))^2}{4(\tau - s)}} ds \\
	&- \frac{g(\tau)}{\sqrt{\pi}}   +\int_0^\tau \frac{ g(s) e^{-\frac{(y(\tau) - y(s))^2}{4(\tau - s)}}  -g(\tau)}{2\sqrt{\pi (\tau - s)^3}} ds
	-\int_0^\tau \frac{ g(s) (y(\tau ) - y(s))}{4\sqrt{\pi (\tau - s)^5}} e^{-\frac{(y(\tau) - y(s))^2}{4(\tau - s)}}  ds \\
	&+\int_0^\tau \int_{y(s)}^{\infty}\frac{\lambda(s, \xi)}{4\sqrt{\pi (\tau - s)^3}} \left[- (y(\tau) -\xi)e^{-\frac{(\xi - y(\tau))^2}{4(\tau-s)}} + (\xi - y(\tau)) e^{-\frac{(\xi -y(\tau))^2}{4(\tau-s)}} \right]  d\xi ds. \nonumber
\end{align*}
This expression can be further simplified by using the identity
\begin{align*}
	\frac{e^{-\frac{(y(\tau) - y(s))^2}{4(\tau - s)}} }{2\sqrt{(\tau -s)^3}} \left(1 + y'(s) (y(\tau) -y(s)) - \frac{(y(\tau) - y(s))^2}{2(\tau -s)} \right) ds =
	d\left(  \frac{e^{-\frac{(y(\tau) - y(s))^2}{4(\tau - s)}}- 1}{\sqrt{\tau  - s}}\right) + \frac{ds}{2 \sqrt{(\tau - s)^3}}.
\end{align*}
Indeed,
\begin{align} \label{Psi_tmp}
	\Psi(\tau, y(\tau)) &= \int_{y(0)}^\infty f(\xi ) \frac{\xi - y(\tau)}{2\sqrt{\pi \tau ^3}} e^{-\frac{(\xi - y(\tau))^2}{4\tau}} d\xi +
	\int_0^\tau \frac{\Psi(s,y(s)) (y(\tau) - y(s))}{2\sqrt{\pi (\tau - s)^3}}  e^{-\frac{(y(\tau) - y(s))^2}{4(\tau - s)}}  ds \\
	&-\frac{g(\tau)}{\sqrt{\pi \tau}} + \int_0^\tau  \frac{g(s) - g(\tau)}{2\sqrt{\pi  (\tau - s)^3}} ds  + \int_0^\tau  \frac{g(s)}{ \sqrt{\pi}} d \left(  \frac{e^{-\frac{(y(\tau) - y(s))^2}{4(\tau - s)}}- 1}{\sqrt{\tau  - s}}\right) \nonumber \\
	&+\int_0^\tau \int_{y(s)}^{\infty} \lambda(s, \xi) \frac{\xi - y(\tau)}{2\sqrt{\pi (\tau - s)^3}} e^{-\frac{(\xi - y(\tau))^2}{4(\tau-s)}} d\xi ds. \nonumber
\end{align}
 Integrating by parts in each integral in \eqref{Psi_tmp}
\begin{align*}
	\int_0^\tau  \frac{g(s) - g(\tau)}{2\sqrt{\pi  (\tau - s)^3}} ds &= \lim_{s \to \tau} \frac{g(s) -  g(\tau)}{\sqrt{\pi (\tau - s)}} - \frac{g(0) - g(\tau)}{\sqrt{\pi \tau}} - \int_0^\tau  \frac{g'(s) ds}{ \sqrt{\pi (\tau - s)}} \\
	\int_0^\tau \frac{g(s)}{\sqrt \pi } d\left( \frac{e^{-\frac{(y(\tau) - y(s))^2}{4(\tau - s)}}- 1}{\sqrt{\tau  - s}} \right) &=
	\lim_{s \to \tau}   \frac{g(s)}{\sqrt \pi } \left( \frac{e^{-\frac{(y(\tau) - y(s))^2}{4(\tau - s)}}- 1}{\sqrt{\tau  - s}}\right) - \frac{g(0)}{\sqrt \pi } \left( \frac{e^{-\frac{(y(\tau) - y(0))^2}{4\tau }}- 1}{\sqrt{\tau }} \right) \\
	&-\int_0^\tau \frac{g'(s)}{\sqrt \pi } \left( \frac{e^{-\frac{(y(\tau) - y(s))^2}{4(\tau - s)}}- 1}{\sqrt{\tau  - s}} \right) ds \\
	\int_{y(0)}^\infty f(\xi ) \frac{\xi - y(\tau)}{2\sqrt{\pi \tau ^3}} e^{-\frac{(\xi - y(\tau))^2}{2\tau}} d\xi  &= -
	\int_{y(0)}^\infty \frac{ f(\xi ) }{\sqrt{\pi \tau}} d \left(e^{-\frac{(\xi - y(\tau))^2}{4\tau}} \right) \\
	&=  \frac{f(y(0))}{\sqrt{\pi \tau}} e^{-\frac{(y(0) - y(\tau))^2}{4\tau}}+\int_{y(0)}^\infty \frac{ f'(\xi ) }{\sqrt{\pi \tau}} e^{-\frac{(\xi - y(\tau))^2}{4\tau}}  d \xi,
\end{align*}
\noindent having in mind that $f(y(0)) = g(0)$ and combining all these terms together yields
\begin{align}
	\Psi(\tau, y(\tau)) &= \int_{y(0)}^\infty f'(\xi ) \frac{e^{-\frac{(\xi - y(\tau))^2}{4\tau}}}{\sqrt{\pi \tau}}  d\xi +
	\int_0^\tau \Psi(s,y(s)) \frac{ (y(\tau) - y(s))  e^{-\frac{(y(\tau) - y(s))^2}{4(\tau - s)}}  }{2\sqrt{\pi (\tau - s)^3}} ds
	-\int_0^\tau g'(s)  \frac{e^{-\frac{(y(\tau) - y(s))^2}{4(\tau - s)}} }{\sqrt {\pi (\tau - s) }}   ds  \nonumber \\
	&+\int_0^\tau \int_{y(s)}^{\infty} \lambda(s, \xi) \frac{\xi - y(\tau)}{2\sqrt{\pi (\tau - s)^3}} e^{-\frac{(\xi - y(\tau))^2}{4(\tau-s)}} d\xi ds.
\end{align}

\end{document}